\documentclass[a4paper,11pt]{article}

\usepackage{jinstpub} 
\usepackage{color}
\usepackage{lineno}
\title{A novel mechanical design of a bolometric array for the CROSS double-beta decay experiment}

\author[a]{D.~Auguste,} 
\author[b]{A.S.~Barabash,}
\author[c]{V.~Berest,}
\author[a]{L.~Berg{\'e},}
\author[d,e,f]{J.M.~Calvo-Mozota,}
\author[g,h]{P.~Carniti,}
\author[a]{M.~Chapellier,}
\author[i]{I.~Dafinei,}
\author[j,k]{F.A.~Danevich,}
\author[a,*]{T.~Dixon,}
\author[a]{L.~Dumoulin,}
\author[c]{F.~Ferri,}
\author[a]{A.~Gallas,}
\author[a]{A.~Giuliani,}
\author[g]{C.~Gotti,}
\author[c]{P.~Gras,}
\author[l]{A.~Ianni,}
\author[a]{L.~Imbert,}
\author[c]{H.~Khalife,}
\author[j]{V.V.~Kobychev,}
\author[b]{S.I.~Konovalov,}
\author[a]{P.~Loaiza,}
\author[a]{P.~de~Marcillac,}
\author[a]{S.~Marnieros,}
\author[a]{C.A.~Marrache-Kikuchi,}
\author[m,n]{M.~Martinez,}
\author[c]{C.~Nones,}
\author[a]{E.~Olivieri,}
\author[m]{A.~Ortiz de Sol\'orzano,}
\author[a]{Y.~Peinaud,}
\author[g]{G.~Pessina,}
\author[a]{D.V.~Poda,}
\author[a]{Ph.~Rosier,}
\author[a]{J.A.~Scarpaci,}
\author[j,l]{V.I.~Tretyak,}
\author[b]{V.I.~Umatov,}
\author[j]{M.M.~Zarytskyy,}
\author[c]{and A.~Zolotarova}

\affiliation[a]{Universit\'e Paris-Saclay, CNRS/IN2P3, IJCLab, 91405 Orsay, France}
\affiliation[b]{National Research Center Kurchatov Institute, Kurchatov Complex of Theoretical and Experimental Physics, 117218 Moscow, Russia}
\affiliation[c]{IRFU, CEA, Université Paris-Saclay, 91191 Saclay, France}
\affiliation[d]{Laboratorio Subterr\'aneo de Canfranc, 22880 Canfranc-Estaci\'on, Spain}
\affiliation[e]{Escuela Superior de Ingenier\'ia, Ciencia y Tecnolog\'ia, Universidad Internacional de Valencia -- VIU, 46002 Valencia, Spain}
\affiliation[f]{Escuela Superior de Ingenier\'ia y Tecnolog\'ia, Universidad Internacional de La Rioja, 26006 Logro\~no, Spain}
\affiliation[g]{INFN Sezione di Milano-Bicocca, I-20126 Milan, Italy}
\affiliation[h]{Universit\`{a} di Milano-Bicocca, Milano I-20126, Italy} 
\affiliation[i]{INFN Sezione di Roma, I-00185 Rome, Italy}
\affiliation[j]{Institute for Nuclear Research of NASU, 03028 Kyiv, Ukraine}
\affiliation[k]{INFN Sezione di Roma Tor Vergata, I-00133 Rome, Italy}
\affiliation[l]{INFN Laboratori Nazionali del Gran Sasso, I-67100 Assergi (AQ), Italy}
\affiliation[m]{Centro de Astropart\'iculas y F\'isica de Altas Energ\'ias, Universidad de Zaragoza, 50009 Zaragoza, Spain}
\affiliation[n]{ARAID Fundaci\'on Agencia Aragonesa para la Investigaci\'on y el Desarrollo, 50018  Zaragoza, Spain}
\affiliation[*]{Currently at University College London, London, UK}

\emailAdd{andrea.giuliani@ijclab.in2p3.fr}
\emailAdd{denys.poda@ijclab.in2p3.fr}

\abstract{  
The CROSS experiment will search for neutrinoless double-beta decay using a specific mechanical structure to hold thermal detectors. The design of the structure was tuned to minimize the background contribution, keeping an optimal detector performance. A single module of the structure holds two scintillating bolometers (with a crystal size of 45 $\times$ 45 $\times$ 45~mm and a Ge slab facing the crystal's upper side) in the Cu frame, allowing for a modular construction of a large-scale array. Two designs are released: the initial \emph{Thick} version contains around 15\% of Cu over the crystal mass (lithium molybdate, LMO), while this ratio is reduced to $\sim$6\% in a finer (\emph{Slim}) design. Both designs were tested extensively at aboveground (IJCLab, France) and underground (LSC, Spain) laboratories. In particular, at LSC we used a pulse-tube-based CROSS facility to operate a 6-crystal array of LMOs enriched/depleted in $^{100}$Mo. The tested LMOs show high spectrometric performance in both designs; notably, the measured energy resolution is 5--7~keV FWHM at 2615~keV $\gamma$s, nearby the Q-value of $^{100}$Mo (3034 keV). Due to the absence of a reflective cavity around LMOs, a low scintillation signal is detected by Ge bolometers: $\sim$0.3~keV (150 photons) for 1-MeV $\gamma$($\beta$) LMO-event. Despite that, an acceptable separation between $\alpha$ and $\gamma$($\beta$) events is achieved with most devices. The highest efficiency is reached with light detectors in the \emph{Thick} design thanks to a lower baseline noise width (0.05--0.09~keV RMS) when compared to that obtained in the \emph{Slim} version (0.10--0.35~keV RMS). Given the pivotal role of bolometric photodetectors for particle identification and random coincidences rejection, we will use the structure here described with upgraded light detectors, featuring thermal signal amplification via the Neganov-Trofimov-Luke effect, as also demonstrated in the present work. 
}


\arxivnumber{2405.18980} % only if you have one

\begin{document}
\maketitle
\flushbottom

%==============================================================================
\section{Introduction}
\label{sec:intro}

The technology of low-temperature detectors \cite{Enss:2005, Pirro:2017, Biassoni:2020} --- devices operating close to 0~K and called as cryogenic calorimeters or bolometers --- is a powerful method for studying lepton number violation through the investigation of a rare nuclear transition known as neutrinoless double-beta ($0\nu\beta\beta$) decay, involving the quasi-simultaneous conversion of two neutrons to two protons with the emission of only two electrons~\cite{GomezCadenas:2023,Agostini:2023,Workman:2022}. A simple scheme of a bolometric detector for a $0\nu\beta\beta$ decay search is represented by a crystal absorber (characterized by a low heat capacity at low temperatures) containing an isotope of interest (a $0\nu\beta\beta$ candidate) and having a weak thermal contact to a heat sink kept at a stable temperature of $O$(10 mK), provided by a special cryostat (a dilution refrigerator), and instrumented with a phonon sensor (in contact with the absorber) to detect particle interactions and read out signals through a dedicated electronic chain (using room- or cold-temperature electronics). A sensor technology typically exploits the temperature dependence of a material on either resistance or magnetization or kinetic inductance. The one of the most robust technologies of phonon sensors, widely used in bolometric $0\nu\beta\beta$ decay searches, is neutron transmutation doped (NTD) Ge thermistors \cite{Haller:1994} readout with room-temperature electronics. Other types of phonon sensors like superconducting transition-edge sensors (TESs), metallic magnetic calorimeters (MMCs), and microwave kinetic inductance detectors (MKIDs) are mainly used in R\&Ds for this application \cite{Poda:2017} (currently, MMCs are also used in $0\nu\beta\beta$ searches \cite{Zolotarova:2020}).

CUORE (Cryogenic Underground Observatory of Rare Events) is the most sensitive bolometric experiment to date, with nearly 1000 detectors based on tellurium dioxide (TeO$_2$) crystals, each measuring 50 $\times$ 50 $\times$ 50~mm and totaling about 750 kg  \cite{Adams:2022a}. CUORE, which uses pure thermal detectors with only heat readout, aims to detect $0\nu\beta\beta$ decay of $^{130}$Te at the Gran Sasso underground laboratory in Italy. CUORE's single readout precludes particle identification capability, thereby leading to a background in the region of interest (ROI) dominated by $\alpha$ particle interactions at a level of approximately $\sim$10$^{-2}$ counts/keV/kg/yr, due to surface radioactive contamination \cite{Adams:2022, Adams:2024b}.

Future large-scale bolometric experiments are considering a double-readout approach, combining heat and light measurements in each event \cite{Artusa:2014wnl, Alenkov:2015, Wang:2015raa, CUPIDInterestGroup:2019inu, Zolotarova:2020, Biassoni:2020}; the registration of photons emitted by the absorber crystal as a result of particle interactions is typically done by another low-temperature detector (e.g. made of a thin Ge or Si wafer) \cite{Poda:2021}. Promising materials for this application are again TeO$_2$ crystals, exploiting the predominant Cherenkov light emitted by $\beta$ particles to reject the $\alpha$ background \cite{Tabarelli:2010,Poda:2017}, and lithium molybdate (Li$_{2}$MoO$_4$, LMO) crystals. Thanks to the scintillation properties of Li$_{2}$MoO$_4$, allowing for particle identification through light-yield difference between $\beta$ and $\alpha$ particles, and to the higher $Q$-value ($Q_{\beta\beta}$) of the $^{100}$Mo $\beta\beta$ transition compared to $^{130}$Te (3034~keV vs. 2528~keV), the background in the ROI is expected to be reduced by two orders of magnitude with respect to the CUORE level, down to $\sim$10$^{-4}$ counts/keV/kg/yr \cite{CUPID_Bkg_Model:2024}. To go even beyond this objective, it is important to reduce also the contribution from $\beta$ particles emitted by surface radioactive impurities \cite{Artusa:2014wnl}.

In this context, the CROSS (Cryogenic Rare-event Observatory with Surface Sensitivity) project aims to develop a demonstrator using metal-coated bolometers based on Li$_{2}$MoO$_4$ and TeO$_2$ crystals for identification of near-surface interactions through pulse-shape discrimination of heat signals  \cite{Bandac:2020,Khalife:2020,Khalife:2020a,Zolotarova:2020,Bandac:2021,Khalife:2021} for a further background reduction in next-generation bolometric $0\nu\beta\beta$ experiments. The CROSS demonstrator plans to combine redundantly pulse-shape discrimination induced by metal-film coating with the double heat-light readout strategy. Following purification and crystallization protocols, developed by LUMINEU \cite{Grigorieva:2017,Armengaud:2017} and applied for CUPID-Mo \cite{Armengaud:2020a,Augier:2022}, 32 Li$_{2}${}$^{100}$MoO$_4$ crystals with 97.7(3)\% $^{100}$Mo enrichment have been produced for the CROSS experiment \cite{Armatol:2021b}. Unlike the pioneering scintillating bolometer experiments CUPID-0 \cite{Azzolini:2018tum} and CUPID-Mo \cite{Armengaud:2020a}, carried out with cylindrical crystals (Zn$^{82}$Se and Li$_{2}${}$^{100}$MoO$_4$ respectively), CROSS crystals are cubic as in CUORE with the aim to maximize packaging inside the experimental volume, thus increasing the detector mass and the efficiency of multi-crystal coincidences. CROSS Li$_{2}$MoO$_4$ cubic crystals have sides measuring 45~mm and a mass of 0.28 kg \cite{Armatol:2021b}.

Considering background studies from previous experiments like CUORE \cite{Alduino:2017, Adams:2024b}, CUPID-0 \cite{Azzolini:2019nmi} and CUPID-Mo \cite{Augier:2023a}, CROSS is designed with minimal passive materials and no reflector around crystal scintillators to mitigate surface radioactivity impact and enhance event coincidences. No detector structure of the above mentioned experiments can be directly applied for CROSS, which will adopt a new design tailored for reduced passive elements, simplified assembly, optimal scintillation detection, and overall improved detector performance. This new design is detailed in the present article. Notably, the upcoming next-generation CUPID (CUORE Upgrade with Particle IDentification) experiment \cite{CUPIDInterestGroup:2019inu} will also adopt the scintillating bolometer technology based on Li$_{2}${}$^{100}$MoO$_4$ crystals with the same shape and size as in CROSS. For this reason, the CROSS crystals have already been used for joint CROSS/CUPID R\&D activities \cite{Armatol:2021a,Armatol:2021b,CrossCupidTower:2023a}, and the study performed here on the CROSS detector holders may be relevant also for the CUPID experiment.

%==============================================================================
\section{R\&D on a detector mechanical structure}

%-------------------------------------------------------------------------------------
\subsection{General requirements for the design}

In order to comply with the demands of the CROSS experiment \cite{Bandac:2020}, we carried out an R\&D on a new mechanical structure for a scintillating bolometer array, keeping in mind the following considerations for the conceptual design:
\begin{itemize}

   \item Active detector components will consist of cubic LMOs (45 $\times$ 45 $\times$ 45~mm) and square-shaped Ge wafers (45~mm side, thickness 0.3 mm);
   
   \item The temperature signals will be read out by NTD Ge thermistors glued at the LMO crystals and at the Ge wafers as in CUORE (and its predecessors), CUPID-0, CUPID-Mo, CROSS, and foreseen in CUPID;
   
    \item The structure will consist of towers with 2 crystals per floor (up to 8 floors to fit the experimental volume of the CROSS cryostat);
    
    \item The Cu-to-LMO mass ratio will be kept as low as possible to minimize radioactivity from bulk and surface of close elements. (For example, this ratio for the detector structures of CUPID-Mo \cite{Armengaud:2020a}, CUPID-0 \cite{Azzolini:2018tum}, and CUORE-0/CUORE \cite{Alduino2016b, Alduino:2018} experiments is around 160\%, 40\%, and 18\%, respectively. The CUORE structure is rather fine but it does not contain bolometric photodetectors and their implementation can increase the Cu-to-LMO mass ratio \cite{Armatol:2021a}.);

    \item Despite poor light output of LMOs, no reflective foil will be used around each crystal to avoid the presence of a large-area source of surface radioactivity and to improve coincidences between detectors (the impact of the presence / absence of a foil is investigated in CUPID-0 \cite{Azzolini:2023});
    
    \item Each LMO cube will have a direct heat sinking through supporting elements made of polytetrafluoroethylene (PTFE) or other plastic material;

    \item Each detector module will be independent of the others in terms of the thermal network;
    
    \item A specific Cu holder for the Ge wafer will be avoided: the wafer will be kept by the same supporting elements that fix the crystal from the top. The gravity resting solution (the full-contact coupling) for the Ge wafer, proposed in \cite{Barucci:2019} and demonstrated with a TeO$_2$ bolometer, needs to be studied with LMOs too;
    
    \item The structure should allow us to bond the NTD thermistors to read them out after the full assembly of a tower, implying that the bonding pads of the thermistors must lie on a vertical plane and must be accessible laterally by the bonding machine.

\end{itemize}

In the next sections, we overview the most important R\&D steps (realized over a 2-year period since the end of 2020) towards the final design of a new mechanical structure.

%-------------------------------------------------------------------------------------
\subsection{Concept of \emph{Thick} and \emph{Slim} structures}

\begin{figure}
\centering
\includegraphics[width=0.95\textwidth]{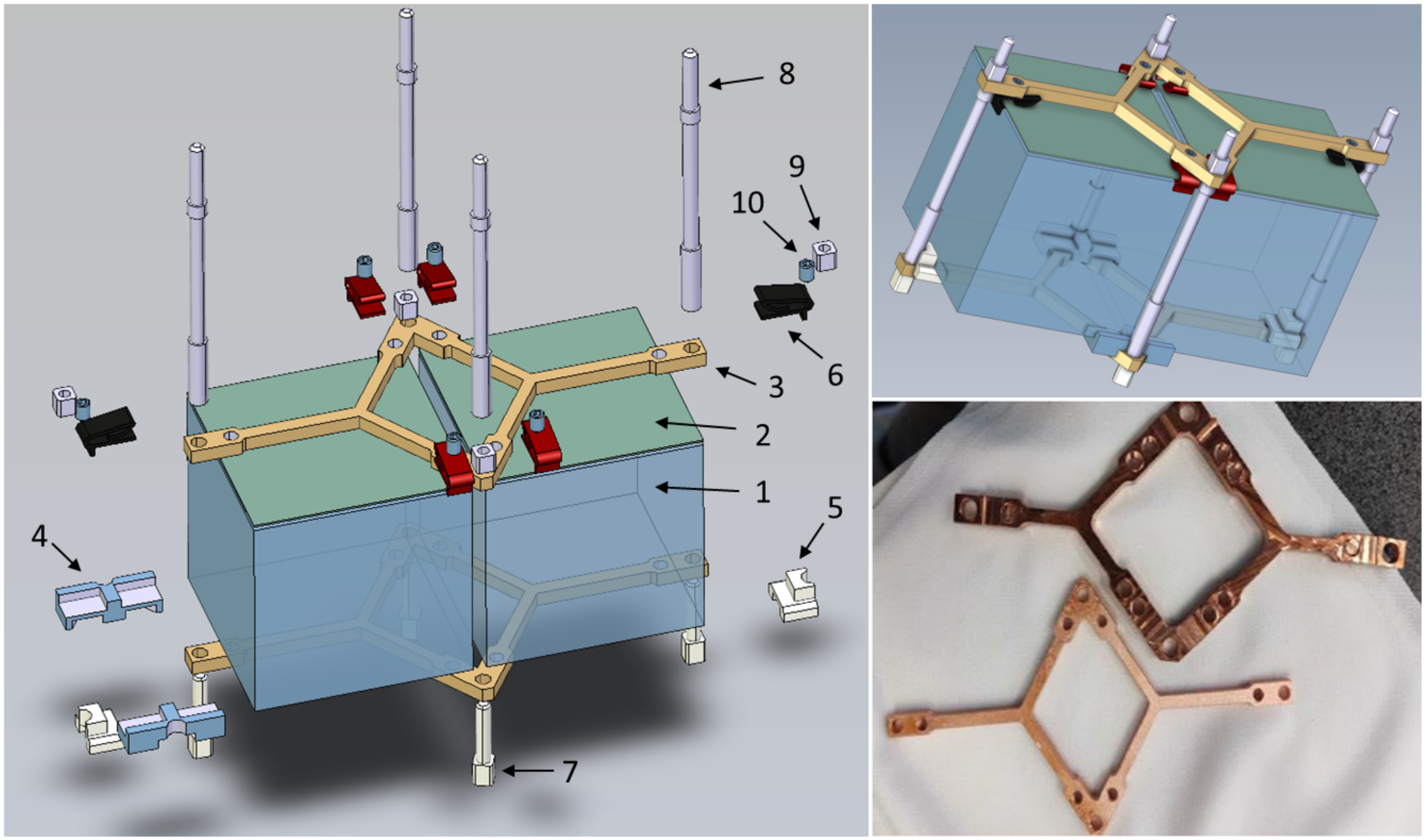}
\caption{Rendering of a two-crystal module with a novel mechanical structure for the CROSS experiment: the constructing components (Left) and a mounted module (Right, top) are shown. 
The CROSS detector structure is made of the following elements: 
(1) two cubic crystals with a 45-mm side; 
(2) two square-shaped Ge wafers (to cover 45 $\times$ 45 mm$^2$ crystal's area on the top);
(3) two Cu frames used on the top and bottom of the construction;
(4, 5) two plastic pieces of two types each to keep the cubic crystals on the bottom Cu frame;
(6) six plastic clamps to keep the Ge wafers;
(7, 8) four Cu rods of two types each to hold the structure;
(9) four Cu nuts to tighten the assembly;
(10) six Cu screws without a head each to press the clamps holding the Ge wafers. 
The Cu frames of the \emph{Thick} and \emph{Slim} designs are presented for comparison (Right, bottom).}
\label{fig:Detector_structure}
\end{figure}

Building on our experience with detector structures featuring single- (section \ref{sec:LD_gravity}) and 4-crystal \cite{Armatol:2021a,CrossCupidTower:2023a} frames, we have developed a new holder concept. This innovative design comprises a double-Y-shaped Cu frame that accommodates two modules of LMO scintillating bolometers (illustrated in figure \ref{fig:Detector_structure}). This structure enables a compact assembly using a comparatively small amount of passive materials. The novel design features three points of support for the detectors and very compact packing of light detectors (LDs). Indeed, the LDs do not have separate copper frames for support, as commonly done in scintillating bolometers, but they are coupled to the crystal through plastic elements. 
Due to such coupling (with the distance of hundreds $\mu$m between the crystal and the wafer), we observe a small cross-talk between the detectors, exhibited as an admixture of a slow component to the fast LD signal (on a level of 0.1\%--1\%, depending on the material, expressing the cross-talk as a ratio between the induced-signal amplitude on the LD and the primary heat signal amplitude). However, this cross-talk does not affect in a significant way the performance of the detectors: only a small correction for light yield evaluation has to be applied. 
The cross-talk issue is mitigated to a negligibly low (if any) contribution using plastic clamps with the thickness of the spacer of about 0.6 mm.

%\begin{figure}[hbt]

\nopagebreak

In the initial (\emph{Thick}) version of the structure, the distance between the Y parts of the frame is larger and the thickness of the frame is higher, comparatively to the re-designed  (\emph{Slim}) structure, as can be seen in figure \ref{fig:Detector_structure} (right, bottom).  Also, we investigated several shapes and materials for spacers between LMO and Ge to avoid both the \emph{full-contact-coupling LD} and \emph{classic LD} approaches; the final design incorporates 3D-printed PLA (polylactic acid) clamps, shown in figure \ref{fig:Detector_structure} (left).  PLA material, used for 3D printing, is commercially available, which allows for a quick adjustment of the pieces' size and a fast production. Before accepting this material for cryogenic measurement, it was tested in terms of radiopurity showing acceptable results for R\&D. At the same time, the radiopurity of PLA material has to be demonstrated at least at the level achieved for PTFE, to be compatible with the background budget of large-scale bolometric experiments. The top frame fixes the two-crystal module, together with four Cu columns, and it serves as a supporting element for the next LMO layer. The size of the structure allows for the formation of a tower consisting of thirteen floors with two-crystal modules per floor, able to fit into a CUORE-like cryostat. The ratio of the Cu elements of the structure to the LMO crystal mass is about 15\% and 6\% for \emph{Thick} and \emph{Slim} designs, respectively. The ratio of the plastic elements to the LMO crystal mass is around 0.4\% for both designs (for comparison, this ratio for a CUORE-like detector structure \cite{Alduino2016b}, holding only crystals, is around 1.5\%). This change in mass was motivated by the goal to reduce as much as possible all passive components, surrounding the detectors, while keeping the rigidity of the structure.  The assembly of the presented structure has to be performed from bottom to top, and each LMO-LD couple is tightened individually using three copper screws which squeeze and deform the three spacers between LMO and Ge wafer.

Also, we switched to a new type of thermal coupling between the absorber and the NTD Ge sensor, i.e. UV-cured glue (Permabond\textregistered) instead of the widely used epoxy gluing (Araldite\textregistered ~bi-component resin), e.g. utilized in the CUORE, CUPID-0, and CUPID-Mo experiments. 

The choice of the components of the detector structure is based on the results of low-temperature tests of several prototypes performed at the cryogenic laboratory of IJCLab. The most representative tests and results are highlighted in the next section.

%-------------------------------------------------------------------------------------
\subsection{Aboveground tests of detector prototypes}

The aboveground location of the cryogenic set-up at IJCLab (Orsay, France) and its comparatively modest shielding against environmental radiation (see in section \ref{sec:facility_ijclab}), are not optimal for the operation of large-volume (e.g. $\sim$100~cm$^3$) thermal detectors due to a high muon- and $\gamma$-induced background counting rate which spoils the detector performance. Despite of that, this set-up offered rapid tests of detector prototypes providing important inputs on the detectors' behavior with respect to the design of the mechanical structure. The richest program of low-temperature studies of detectors, held in the new mechanical structure, was realized with the very first prototype of the \emph{Thick} design. This investigation together with the test of the \emph{full-contact-coupling LD} approach are thus described in detail, while other tests at IJCLab are briefly overviewed.

%-------------------------------------------------------------------------------------
\subsubsection{Aboveground cryogenic facility}
\label{sec:facility_ijclab}

The facility exploits a custom-made $^3$He/$^4$He dilution refrigerator with a pulse-tube-based pre-cooling down to 4 K \cite{Mancuso:2014a}. The cryostat features a comparatively large experimental volume (around 5 liters) and a reasonably fast cooling down to millikelvin temperatures (the base temperature is around 10 mK). The outer vacuum chamber of the cryostat is surrounded by a 10-cm-thick lead shield, reducing the background counting rate and mitigating pile-up issues typical for large-volume thermal detectors. The detectors installed inside the facility are connected to a floating plate, which is suspended from the mixing chamber --- the coldest part of the refrigerator --- to reduce vibrations induced by the pulse tube. 

The set-up exploits 6-channel room-temperature voltage-sensitive DC pre-amplifiers, a predecessor of the CUORE front-end electronics \cite{Arnaboldi:2002}. An anti-aliasing Bessel-Thomson filter is also used in the readout. Continuous stream data are acquired with a multi-channel 16-bit analog-to-digital board. Additional 6-channel wiring in the cryostat is used to provide electrical connection to either heating elements or electrodes of bolometers (presented below). In addition, a set of optic fibres are installed inside the cryostat to deliver, in a controlled manner, photons from room-temperature LED to bolometric detectors.

%-------------------------------------------------------------------------------------
\subsubsection{Study of the \emph{full-contact-coupling LD} approach}
\label{sec:LD_gravity}

The \emph{full-contact-coupling LD} approach has been proposed in \cite{Barucci:2019} as a way to keep a Ge wafer on a crystal top surface, avoiding any additional supporting elements. A prototype constructed from a CUORE-like TeO$_2$ crystal (with a size of about 50 $\times$ 50 $\times$ 50~mm) and a Ge-based light detector ($\oslash$44 $\times$ 0.18 mm), was tested at $\sim$11~mK base temperature in a liquid-helium-bath cryostat at the Gran Sasso underground laboratory, in Italy  \cite{Barucci:2019}. The authors reported excellent performance of the LD, such as 3.9~$\mu$V/keV sensitivity (a corresponding NTD resistance is 1.5~M$\Omega$) and 20~eV RMS on the filtered baseline noise. Moreover, the \emph{full-contact-coupling LD} approach allowed to gain about 50\% in the detected light signal ($\sim$150~eV LD signal is measured per 2.6~MeV energy deposited by $\gamma$'s in the TeO$_2$ crystal) \cite{Barucci:2019}. These experimental results together with simulations for LMOs \cite{Armatol:2021a} motivated us to perform a similar study with an LMO crystal\footnote{Recently, a test with a slightly larger octagonal-shaped LD, in the full-contact-coupling to a 45-mm side cubic LMO, has been presented in \cite{Alfonso:2022}, reporting a similar conclusion concerning the usability of the approach for LMOs.}. 

\begin{figure}
\centering
\includegraphics[width=0.95\textwidth]{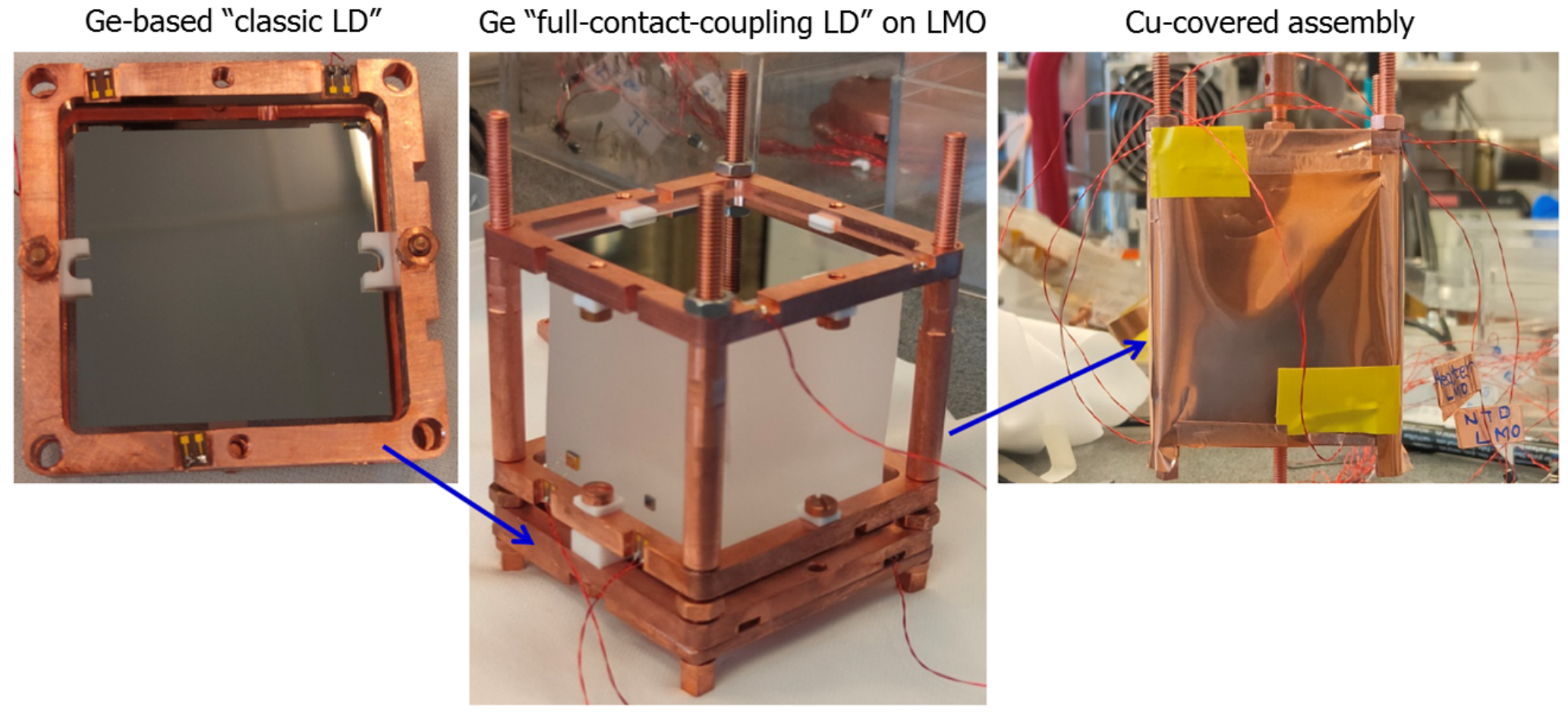}
\caption{An assembly of a 45-mm-side cubic LMO bolometer with two identical Ge light detectors (shaped to 45 $\times$ 45~mm$^2$ active area), one is resting on the crystal top (\emph{full-contact-coupling LD}) and the second device is mounted in the independent Cu holder (\emph{classic LD}) at the LMO bottom, 5~mm far from the crystal surface. The full detector array was covered by a Cu foil, to improve the shielding against the infrared radiation, and connected to a spring-suspended Cu plate inside a pulse-tube cryostat.}
\label{fig:LD_gravity_Photo}
\end{figure}

For a \emph{full-contact-coupling LD} test, we used one crystal scintillator from the CROSS batch and two identical Ge wafers (45 $\times$ 45 $\times$ 0.3~mm), which had no thin-film coating capable of enhancing the light collection efficiency by reducing the reflection from the Ge surface \cite{Mancuso:2014,Azzolini:2018tum}. The absorbers were equipped with NTD Ge thermistors, glued using Araldite\textregistered~Rapid epoxy. All NTD thermistors used (here and in the below-presented experiments) were produced from Ge wafers with similar irradiation parameters, yielding to similar resistance-vs-temperature characteristics, which can be approximated as $R(T) = R_0 \cdot e^{(T_0/T )^{0.5}}$ with the irradiation-dependent parameter $T_0$ $\sim$ 3.7~K and the sensor-size-dependent parameter $R_0$ $\sim$ 1~$\Omega$ (for 3 $\times$ 3 $\times$ 1~mm NTDs). 
A Si chip with a Si:P meander micro-fabricated on it \cite{Andreotti:2012} was also glued onto the LMO surface, serving as a heater for detector's response test and stabilization. The LMO crystal was mounted between two square-shaped Cu frames, connected by four Cu columns, using PTFE supporting elements. One LD was constructed using an identical Cu frame and two PTFE clamps. The second LD was placed directly on the LMO top and constrained with PTFE pieces. The full assembly is shown in figure~\ref{fig:LD_gravity_Photo}. The detector module was covered by a Cu tape, to improve the infrared radiation shielding, and then installed inside the pulse-tube-based cryostat described above (section \ref{sec:facility_ijclab}).

We carried out measurements at 15~mK, regulated at the detector plate. For calibration purpose, we deployed a $^{232}$Th source inside the Pb shield. The measurements were affected by sub-optimal noise conditions as well as a high counting rate of the LMO bolometer: however, this impacted the operation of the detectors and their performance, but not the main goal of the study. Data were acquired with the sampling rate of 5 kS/s for all channels. 

The collected data were processed offline using a MATLAB-based tool \cite{Mancuso:2016}, which exploits a matched filter (the Gatti-Manfredi optimum filter \cite{Gatti:1986}) to maximise the signal-to-noise ($S/N$) ratio and to get the information about the signal amplitude (i.e. energy) and several pulse-shape parameters of each triggered event. To construct the transfer function of the optimum filter, we used an average pulse (represented by tens of individual high-energy signals) and an average noise (10000 baseline noise traces) of each channel, extracted from the data. 

The LMO channel was calibrated with 2615~keV $\gamma$ quanta of $^{208}$Tl, emitted from the $^{232}$Th source, while the energy scale of both LDs was determined by X-rays of Mo, induced in the LMO crystal by the impinging radiation, e.g. as done in the CUPID-Mo \cite{Augier:2022}. 
The achieved detector performance are summarized in table~\ref{tab:LD_gravity_performance}. The NTDs were biased with relatively high currents, aiming to make the response faster and to reduce sensitivity to thermal fluctuations.  
It is worth noting that the noise of the \emph{full-contact-coupling LD} channel is found to be a factor 7 worse than that of the \emph{classic LD} one. We can explain that by the typically higher vibration noise of a pulse-tube cryostat \cite{Olivieri:2017} than that of a liquid-helium-bath dilution refrigerator, as one used in the first test of the \emph{full-contact-coupling LD} approach \cite{Barucci:2019}.

\begin{table}
\centering
\caption{Performance of an LMO bolometric module with two light detectors: one device is constructed using an independent holder (\emph{classic LD}), while the second LD is kept following the \emph{full-contact-coupling LD} approach, proposed in \cite{Barucci:2019}. For each channel we are reporting the working point of the NTD thermistor (resistance $R_{NTD}$ at a given current $I_{NTD}$), pulse-shape time constants (rise- and decay-time parameters, $\tau_{r}$ and $\tau_{d}$), detector sensitivity represented by a signal voltage amplitude per unit of the deposited energy ($A_s$), RMS baseline width (RMS$_{noise}$).}
\smallskip
\begin{tabular}{lcccccc}
\hline
Channel  & $R_{NTD}$ & $I_{NTD}$ & $\tau_{r}$ & $\tau_{d}$ & $A_s$ & RMS$_{noise}$ \\
~ & (M$\Omega$)  & (nA) & (ms) &  (ms) & (nV/keV) & (keV)  \\
\hline
\hline
LMO         & 0.70 & 4.0 & 14   & 78  & 8 & 22       \\
\hline
\hline
Ge \emph{classic LD}  & 0.53 & 15  & 1.2  & 6.7 & 480 & 0.13   \\
\cline{2-7}
Ge \emph{full-contact-coupling LD}  & 0.60 & 20  & 0.85 & 5.1 & 240 & 0.85   \\
\hline
\end{tabular}
\label{tab:LD_gravity_performance}
\end{table}

\begin{figure}
\centering
\includegraphics[width=0.65\textwidth]{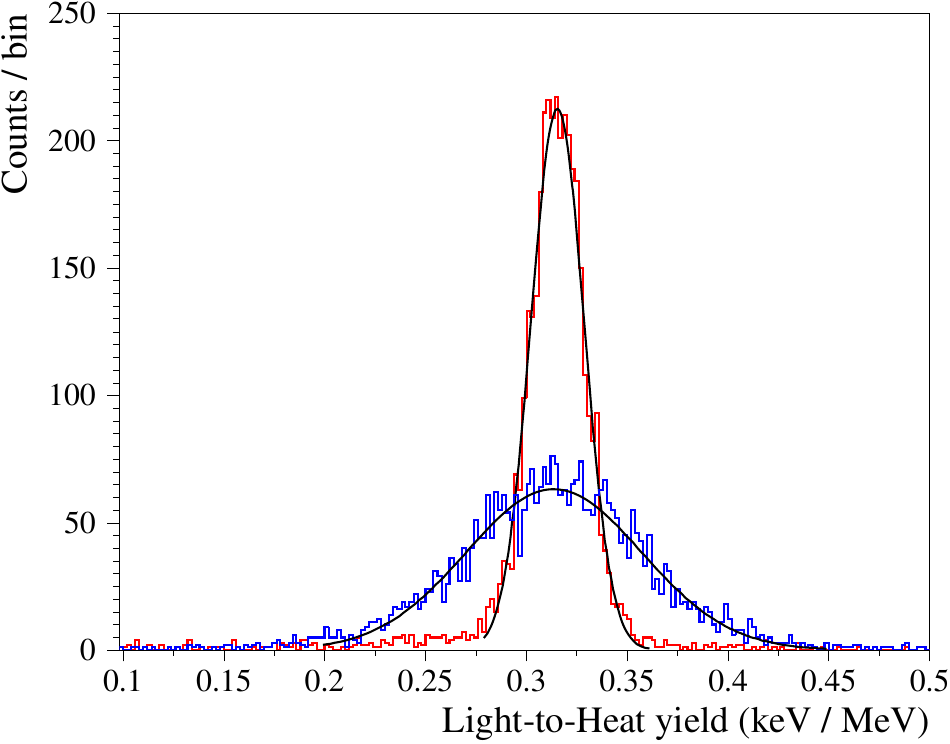}
\caption{Distribution of Light-to-Heat parameter for muons detected in the LMO crystal ($\sim$12--30 MeV energy) in coincidence with scintillation signals measured by the \emph{classic LD} (red) and the \emph{full-contact-coupling LD} (blue) devices. The mean values of the Light-to-Heat parameter are 0.31(4) and 0.31(1) keV/MeV, respectively; the width of each distribution is taken as an uncertainty of the corresponding Light-to-Heat value.}
\label{fig:LD_gravity_LY}
\end{figure}

Using the information about the rising part of the detectors signals and LMO-triggered events, we established coincidences between the LMO bolometer and two LDs, following a method similar to \cite{Piperno:2011}. Due to the rather poor noise of the \emph{full-contact-coupling LD} channel (see in table~\ref{tab:LD_gravity_performance}) as well as a sub-efficient light collection (the absence of an anti-reflective coating of Ge wafers and a highly efficient reflector around the crystal scintillator) together with the low scintillation yield of the LMO material, the scintillation signals induced by $\gamma(\beta)$s in the LMO-bolometer are mostly hidden in the LD noise. Thus, we compare scintillation light energy detected by the LDs for muons passing through the LMO crystal only and depositing there $\sim$12--30 MeV energy (above natural $\alpha$-, $\beta$-, and $\gamma$-radioactivity). For this purpose, it is convenient to use a dimensionless Light-to-Heat ratio defined as the LD-detected scintillation signal (in keV) normalized on the heat energy release (in MeV) measured by the LMO bolometer, which serves as a particle identification parameter (illustrated below). The distributions of the Light-to-Heat parameter for the selected muon events is presented in figure \ref{fig:LD_gravity_LY}. We found no clear evidence of the enhanced light collection for the \emph{full-contact-coupling LD} channel compared to \emph{classic LD} one. Therefore, the \emph{full-contact-coupling LD} resting solution \cite{Barucci:2019} is discarded from the conceptual design of the mechanical structure for the CROSS experiment.

%--------------------------------------------------
\subsubsection{First prototype of the \emph{Thick} structure}
\label{sec:Prototype_Thick}

\begin{figure}
\centering
\includegraphics[width=1.0\textwidth]{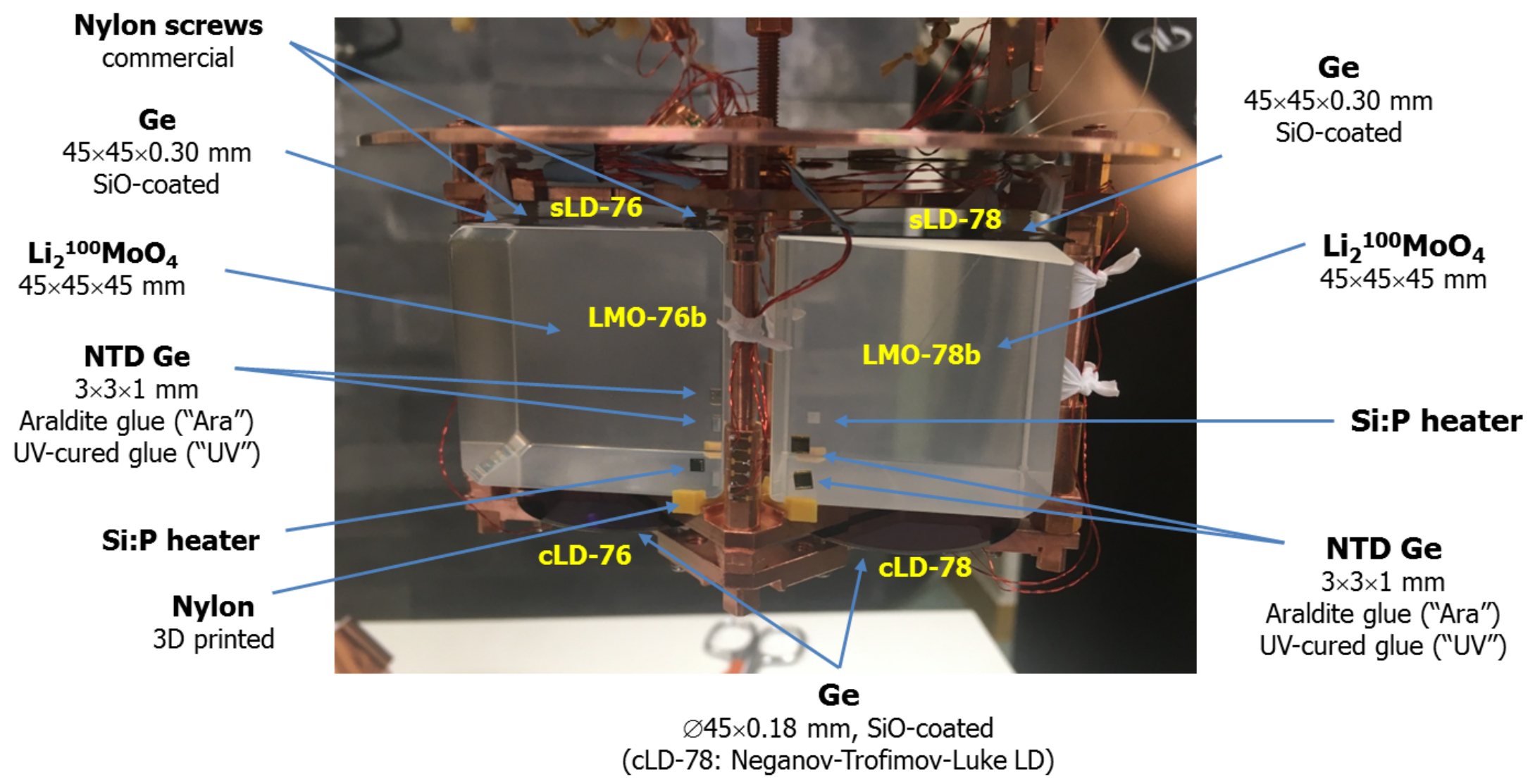}
\caption{Photo of the first prototype of the \emph{Thick} structure detector module design with labeled components.}
\label{fig:Thick_structure_module}
\end{figure}

The first prototype of the \emph{Thick} design was featuring two $^{100}$Mo-enriched LMOs (45 $\times$ 45 $\times$ 45~mm), two square-shaped Ge LDs (45 $\times$ 45 $\times$ 0.30~mm) and two circular-shaped Ge LDs ($\oslash$44 $\times$ 0.18~mm) to collect scintillation light emitted from the top and the bottom sides of the crystals. 
Aiming to improve light collection, all sides of Ge wafers faced to LMOs were SiO-coated, reducing reflection of scintillation photons on the LD surface. One circular-shaped LD has a system of Al electrodes enabling a thermal signal amplification proportional to the voltage applied across two nearby electrodes, exploiting the so-called Neganov-Trofimov-Luke effect, see e.g. \cite{Novati:2019} and references therein. Each Ge wafer was equipped with a small ($\sim$5 mg) NTD Ge thermistor glued with the Araldite\textregistered~Rapid bi-component epoxy. LMOs were instrumented with two NTDs each; one sensor was glued with the epoxy (labeled here as ``Ara''), while another with UV-cured glue (``UV''). The heating element was also coupled to each LMO with the epoxy glue. The detectors were assembled in the Cu frame of the \emph{Thick} design using 3D-printed PLA pieces to decouple the absorbers from the heat bath and Nylon screws to press the top LDs against the crystals. Finally, the electrical contacts to the sensors were provided by $\oslash$25-$\mu$m Au bonding wires; the heaters were bonded with Al wires. We remark that the assembly time of the 2-LMO-crystal module is rather short, which can be broken down into the following processes: NTD gluing with ``UV'' / ``Ara'' type adhesives is approximately  5 min / 1 h; mounting is $\sim$15 min; sensor bonding is $\sim$20 min. 
The detector module connected to the cold stage of the dilution refrigerator is shown in figure \ref{fig:Thick_structure_module}.

\begin{figure}[hbt]
\centering
\includegraphics[width=0.9\textwidth]{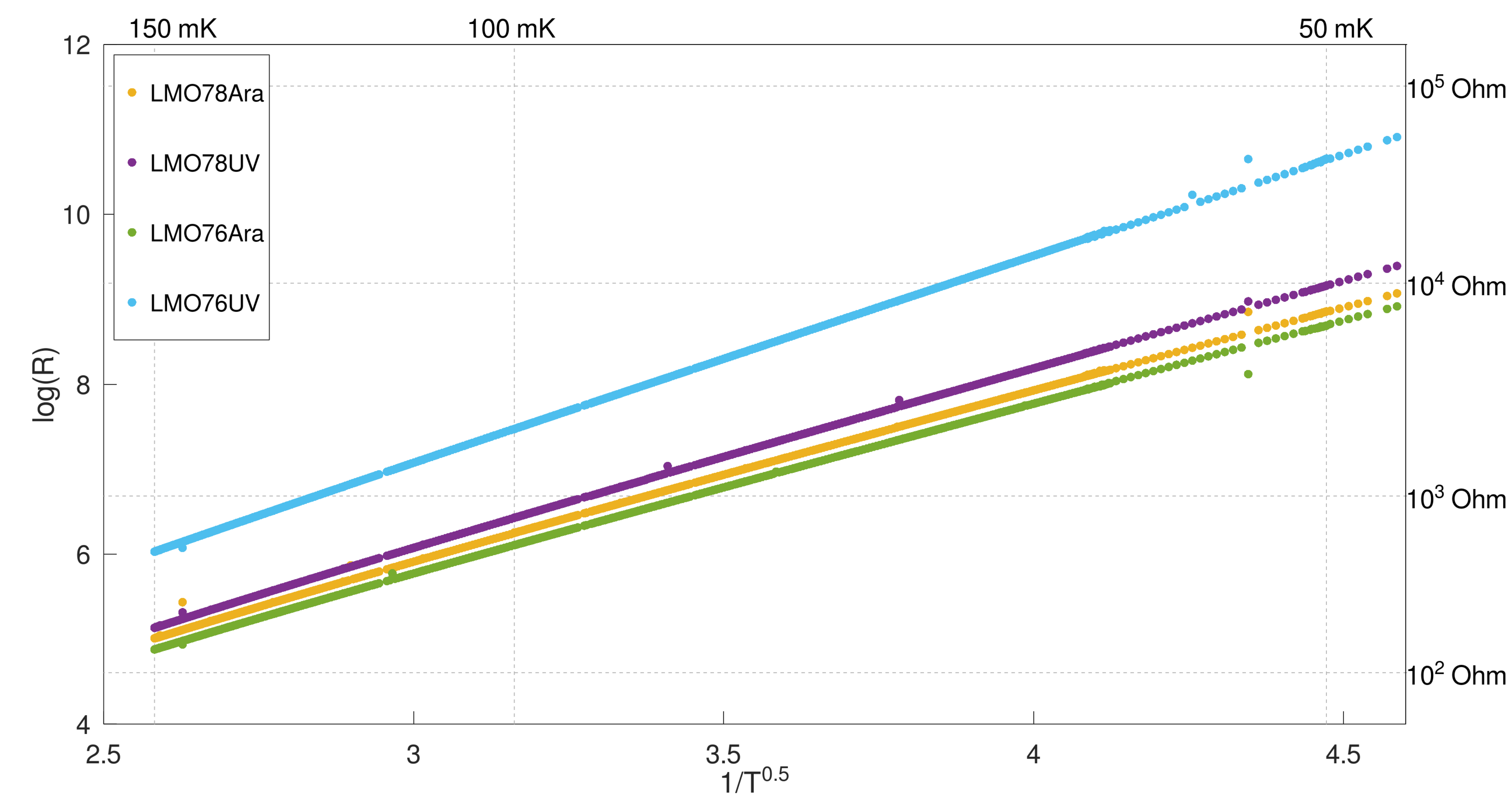}
\caption{Resistance-temperature characterization of NTDs, coupled to LMO crystals with either epoxy (Ara) or UV-cured (UV) glue.}
  \label{fig:RT_run68}
\end{figure}

After reaching base temperature (around 12~mK), we performed a study of the resistance-temperature dependence for all NTDs coupled to LMOs to compare the gluing materials: the coupling can induce extra stress in the sensor, which can translate into a larger value of the $T_0$ parameter characterizing the $R$($T$) behavior of NTDs. With this aim, we collected $R(T)$ data of the LMO sensors below $\sim$100 mK and performed fitting in the 25--80 mK range (see figure~\ref{fig:RT_run68}) to extract the $T_0$ values listed in table \ref{tab:Thick_structure_performance}. The obtained $T_0$ values of ``Ara'' NTDs are almost the same, while both ``UV'' NTDs are characterized by larger $T_0$ values indicating a higher mechanical stress between the sensor and the crystal. These observations are similar to what was reported for Ge light detectors equipped with NTDs and exhibited the impact of the glue-induced stress between the absorber and the sensor \cite{Alfonso:2023}.

Low-temperature measurements of particle-detection capability and data analysis were carried out similarly to the \emph{full-contact-coupling LD} approach test (sec. \ref{sec:LD_gravity}). In particular, the detectors were cooled down to the same temperature (15~mK) to acquire data streams with the same DAQ settings (5 kS/s sampling, 675 Hz Bessel cut-off frequency), which are then processed by the same OF-based tool. Due to the limited number of channels (6) available for the simultaneous readout, two LDs were not operational at the beginning of the experiment, oriented to a direct comparison of NTDs of LMOs. 
This period of measurements was characterized by sub-optimal noise conditions, affecting the performance of the detectors, but not the goal of the study.  
After optimization of noise conditions, we operated only UV-glued NTDs of LMOs and all four LDs to investigate more extensively the new NTD coupling for LMO bolometers and to have a direct comparison of LDs performance.

\begin{figure}[t]
\centering
\includegraphics[width=0.75\textwidth]{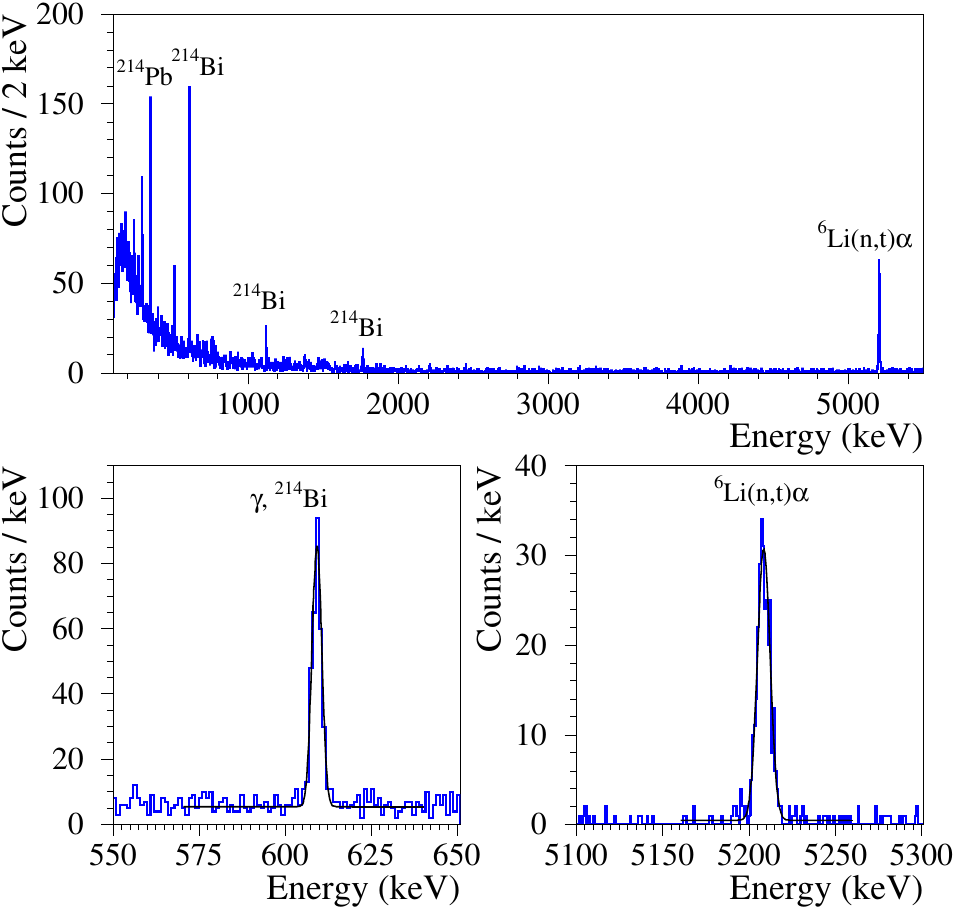}
\caption{(Top) Energy spectrum measured with a 0.28-kg $^{100}$Mo-enriched Li$_2$MoO$_4$ bolometer (LMO-78b) operated in the surface-level laboratory (the measurement time is 15~h). The most intense $\gamma$-ray peaks and the neutron capture peak observed in the spectrum are labeled. (Bottom) Sections of the energy spectrum showing the 609-keV $\gamma$ peak of $^{214}$Bi (Left) and detected products of the $^6$Li(n,t)$\alpha$ reaction ($Q$-value is 4784~keV), induced by environmental neutrons and calibrated on a $\gamma$-energy scale (Right). The fits to the peaks with a Gaussian plus a flat background component are shown by solid lines. The energy resolution of the peaks is 3.3(3) keV FWHM and  8.7(6) keV FWHM, respectively. The energy of $\alpha$+t events, measured in $\gamma$ scale, is shifted with respect to the $Q$-value of the reaction of the thermal neutron capture on $^6$Li due to the thermal quenching of the phonon signal.}
  \label{fig:LMO_Bkg_ijclab}
\end{figure}

\begin{figure}[hbt]
\centering
\includegraphics[width=0.6\textwidth]{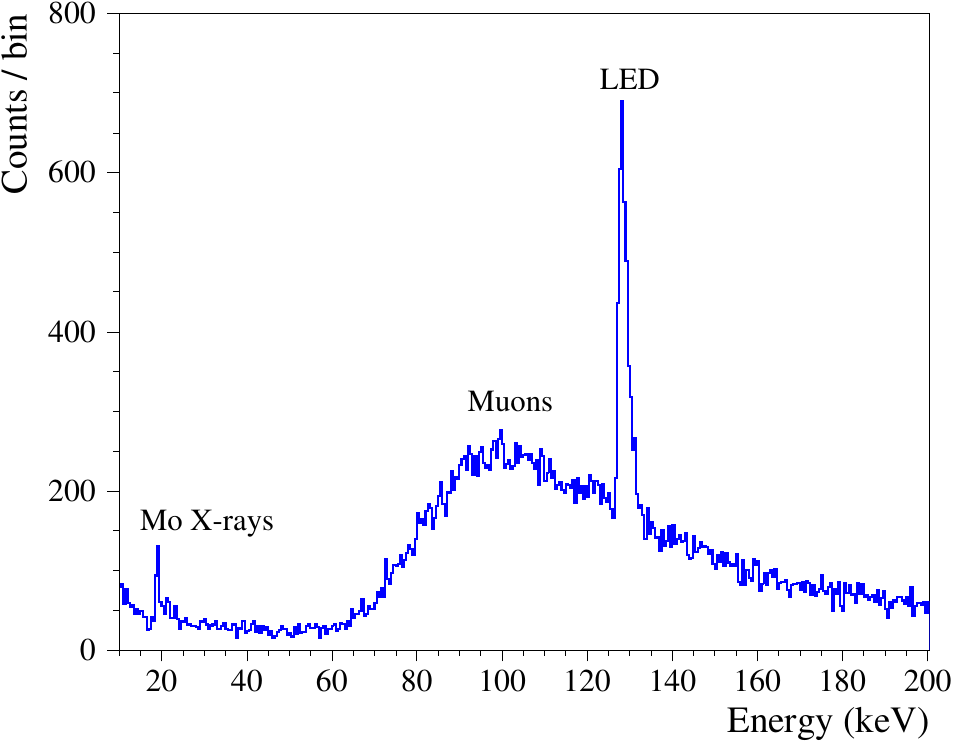}
\caption{Energy spectrum of events detected by the bolometric Ge light detector cLD-76 in 64-h-long measurements in the pulse-tube cryostat. The data exhibit peaks ascribed to the Mo X-rays and LED signals, and a broad distribution of muon-induced events.}
  \label{fig:LD_Bkg_ijclab}
\end{figure}

\begin{figure}[hbt]
\centering
\includegraphics[width=0.49\textwidth]{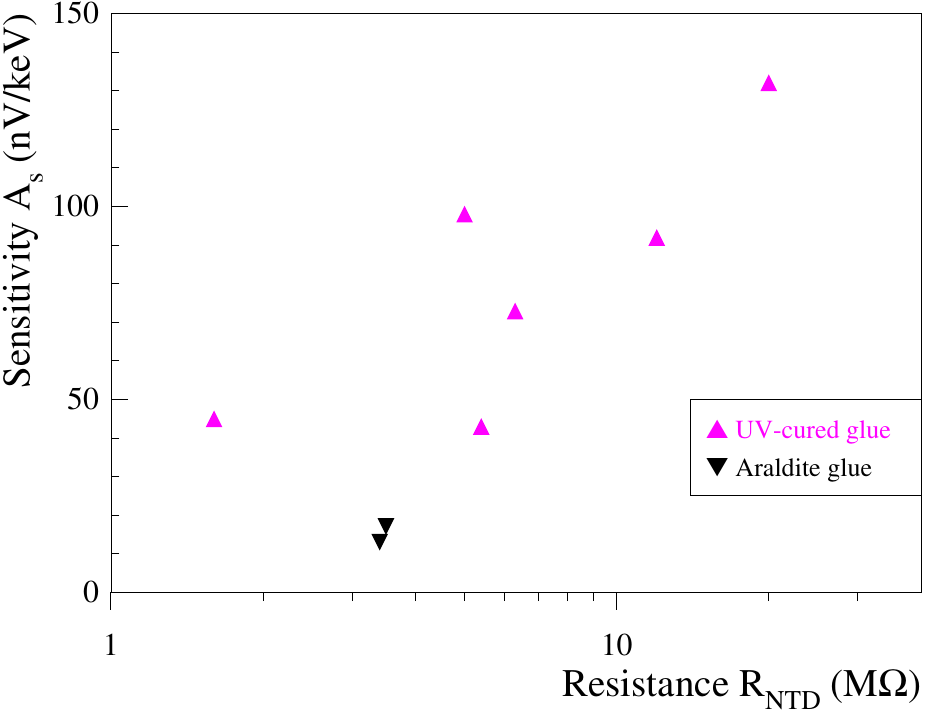}
\includegraphics[width=0.49\textwidth]{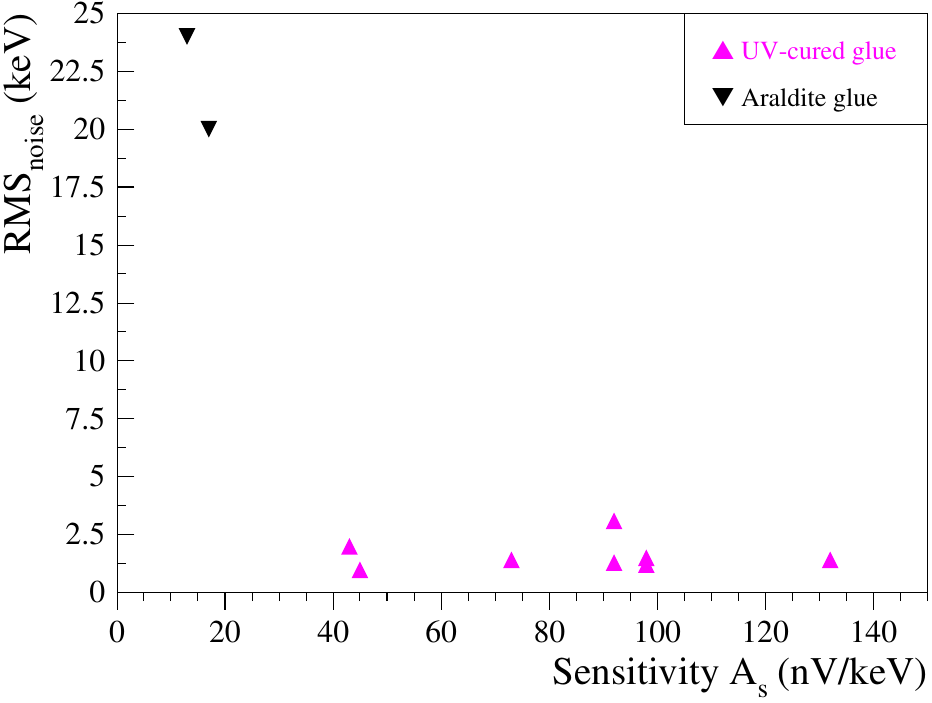}
\caption{Parameters (NTD resistance, sensitivity and noise energy resolution) of two Li$_2$$^{100}$MoO$_4$ bolometers used in the study of two couplings of NTD to LMO crystals.}
  \label{fig:LMO_performance_ijclab}
\end{figure}

\begin{figure}[hbt]
\centering
\includegraphics[width=0.6\textwidth]{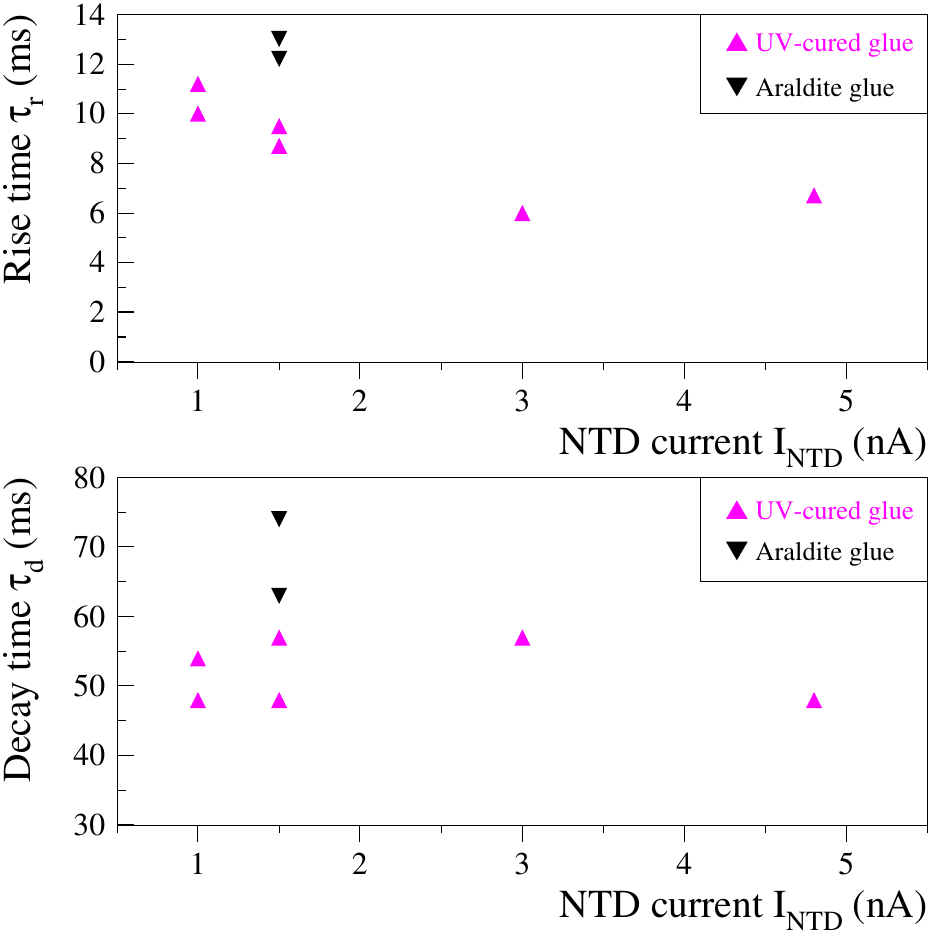}
\caption{Time constants (rise and decay time parameters) of thermal signals of two Li$_2$$^{100}$MoO$_4$ bolometers instrumented with two NTDs each, coupled to the crystals with ether a ``classic'' (epoxy) or a new (UV-cured) type of glue. Rise time shows a clear variation with the applied (relatively high) current, while the decay time remains similar.}
  \label{fig:LMO_time_constants}
\end{figure}

In order to estimate detector performance such as the sensitivity and the energy resolution, we calibrated LMO bolometers with $\gamma$ quanta of $^{214}$Bi (609~keV) and $^{214}$Pb (352~keV) from environmental radioactivity, as illustrated in figure \ref{fig:LMO_Bkg_ijclab}. The LDs calibration has been done with the distribution of muons passing through the Ge wafers, the maximum distribution of which was initially calibrated with Mo X-rays (17.5~keV), as illustrated in figure \ref{fig:LD_Bkg_ijclab}. The mean muon-induced energy is measured as $\sim$100 keV and $\sim$210 keV for circular- and square-shaped Ge LDs, respectively, in agreement with our expectations based on the thickness of the wafer and the energy loss of muons in the material \cite{Novati:2019}. Knowing the energy scale of the detectors we evaluated their performance listed in table \ref{tab:Thick_structure_performance} and illustrated in figures \ref{fig:LMO_performance_ijclab} and \ref{fig:LMO_time_constants}.

%\begin{figure}[hbt]

\begin{table}
\centering
\caption{Best performance of the 2-crystal module of the \emph{Thick} design mechanical structure tested at 15~mK in the pulse-tube cryostat in the aboveground laboratory. In addition to the parameters quoted in table~\ref{tab:LD_gravity_performance}, we report the $T_0$ value for LMO NTDs. For NTDs coupled to LMOs with UV cured glue, we report performance at three working points. The RMS$_{noise}$ value is reported for conditions with optimized noise level but the values labeled with $^*$, which are obtained before the optimization. For the cLD-78 we also report performance achieved thanks to the application of the Neganov-Trofimov-Luke signal amplification (45~V electrode bias), labeled with $^{**}$.}
\smallskip
\begin{tabular}{l|ccccccc}
\hline
Channel  & $T_0$ & $R_{NTD}$ & $I_{NTD}$ & $\tau_{r}$ & $\tau_{d}$ & $A_s$ & RMS$_{noise}$ \\
~ & (K) & (M$\Omega$) & (nA)   & (ms) &  (ms) & (nV/keV) & (keV)  \\
\hline
\hline
LMO-76b Ara      & 3.71 & 3.4 & 1.5 & 13.0  & 74  & 13 & 24$^*$       \\
\cline{2-8}
LMO-76b UV       & 5.74 & 20  & 1.0 & 11.2  & 54  & 132 & 1.4       \\
~                & ~    & 12  & 1.5 & 8.7   & 57  & 92 & 1.3 (3.1$^*$)       \\
~                & ~    & 5.4 & 3.0 & 6.0   & 57  & 43 & 2.0       \\
%\cline{2-8}
\hline
LMO-78b Ara      & 3.70 & 3.5 & 1.5 & 12.2  & 63  & 17 & 20$^*$       \\
\cline{2-8}
LMO-78b UV       & 4.00 & 6.3 & 1.0 & 10.0  & 48  & 73 & 1.4       \\
~                & ~    & 5.0 & 1.5 & 9.5   & 48  & 98 & 1.2 (1.5$^*$)       \\
~                & ~    & 1.6 & 4.8 & 6.7   & 48  & 45 & 0.98       \\
\hline
\hline
sLD-76  & n.a. & 2.5 & 4.8 & 0.98 & 1.8 & 570  & 0.36   \\
\cline{2-8}
cLD-76  & n.a. & 1.7 & 4.8 & 0.87 & 3.9 & 870  & 0.12   \\
\hline
%\cline{2-8}
sLD-78  & n.a. & 1.8 & 4.8 & 1.1  & 3.4 & 740  & 0.12   \\
\cline{2-8}
cLD-78  & n.a. & 1.5 & 4.8 & 0.85 & 3.7 & 1460 & 0.14   \\
~       & ~    & ~   & ~   & ~    & ~   & 15600$^{**}$ & 0.015$^{**}$   \\
\hline
Mean LD (no NTL)  & ~ & 1.5 & 4.8 & 0.94 & 2.9 & 810 & 0.15   \\
\hline
\end{tabular}
\label{tab:Thick_structure_performance}
\end{table}

As seen in table \ref{tab:Thick_structure_performance} and in figure \ref{fig:LMO_performance_ijclab}, in measurements at 15 mK we found that the resistances of UV-glued NTDs (6--20~M$\Omega$) are at least twice higher than those of the epoxy-coupled thermistors ($\sim$3~M$\Omega$) operated at working points with the same current across the NTD (1.5~nA), as expected according to the above reported $R(T)$ study. Furthermore, the NTDs coupled with the UV glue are characterized by faster thermal signals compared to their counterparts with the epoxy coupling, as illustrated in figure \ref{fig:LMO_time_constants}. In particular, the shortest rise time values of UV-glued NTDs achieved in the present work (6--7 ms) are twice shorter than early reported results for the epoxy-glued NTDs on LMOs \cite{Armengaud:2017,CrossCupidTower:2023a}. Such fast response together with a typically high $S/N$ ratio of the LMO heat channel (a factor 100 higher than that of the LD channel) can be crucial for the efficient rejection of pulse pile-up, which could introduce significant background contribution for bolometric experiments \cite{Chernyak:2012,Chernyak:2014}. Also, we see a correlation between NTD resistance (2--20 M$\Omega$) and sensitivity (13--130 nV/keV), as shown in figure \ref{fig:LMO_performance_ijclab} (Left); thus, the more resistive UV-glued NTDs develop notably higher voltage signals compared to the epoxy-glued sensors, leading to about 6 times higher sensitivity for the same applied current. The difference in the baseline noise resolution is even higher: a factor of 10, as seen in figure \ref{fig:LMO_performance_ijclab} (Right). Therefore, during conditions characterized by a higher noise level, LMOs with the epoxy-glued NTDs show a poor baseline resolution as for such type of detectors, while the channels of the UV-glued NTDs have a good baseline resolution. Moreover, the baseline resolution was further improved by a factor of 2 in measurements with optimized noise of the set-up, to a very low level of $\sim$1 keV RMS considering the large detector volume and the high background rate at the sea level. This performance is achieved thanks to both the increased sensitivity and the faster thermal response of NTDs coupled with the UV-cured glue. The low level of the baseline noise allowed us to perform $\gamma$-ray measurements with high energy resolution, as illustrated in figure \ref{fig:LMO_Bkg_ijclab}. This example shows data of an LMO bolometer (LMO-76b UV), characterized by a sensitivity of 45 nV/keV and a baseline noise of 0.98 keV RMS, in which the energy resolution of a 609-keV peak induced by $\gamma$ quanta of $^{214}$Bi is measured as 3.3(3) keV FWHM. Such energy resolution is among the best ever reported values for LMO bolometers in this energy interval \cite{Armengaud:2017}. Moreover, the detector shows good energy resolution at an-order-of-magnitude higher energies, as exhibited by the sharp peak of neutron-capture-induced events (alpha plus triton) measured with 8.7(6) keV FWHM. It is worth emphasizing that these notable results are achieved in an aboveground setup, which is not optimal for the operation of large-size thermal detectors with typically slow response, being affected by a high counting rate induced by cosmic rays and environmental radioactivity.

As it concerns LDs, we biased their NTDs at higher currents, so reducing resistances to a few M$\Omega$ and getting much faster ($\sim$1 ms) rise times of the thermal signals than those of LMOs, thanks to the difference in the size of the absorber and the sensor (i.e. the reduced heat capacity). A further increase of the NTD current would allow us to reach a sub-millisecond rise time, essential for pile-up rejection \cite{Chernyak:2017,CROSSpileup:2023}. Despite the high NTD current, LDs show good sensitivity (0.6--1.5 $\mu$V/keV) for such type of bolometric devices. Three over four LDs are characterized by a low baseline noise, $\sim$0.12 keV RMS, while 3 times worse resolution is measured for a single LD. Moreover, for the LD with Al electrodes, we applied a 45 V bias to amplify the heat signal (exploiting the Neganov-Trofimov-Luke effect), allowing us to improve both sensitivity and baseline noise resolution by a factor of 10.

\begin{figure}[hbt]
\centering
\includegraphics[width=0.49\textwidth]{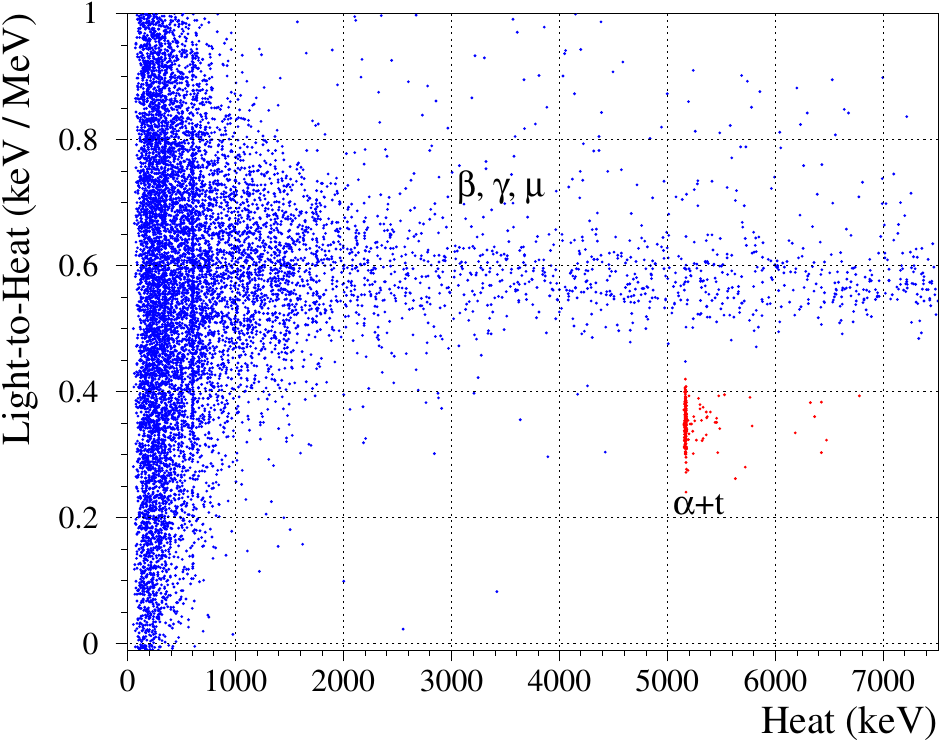}
\includegraphics[width=0.49\textwidth]{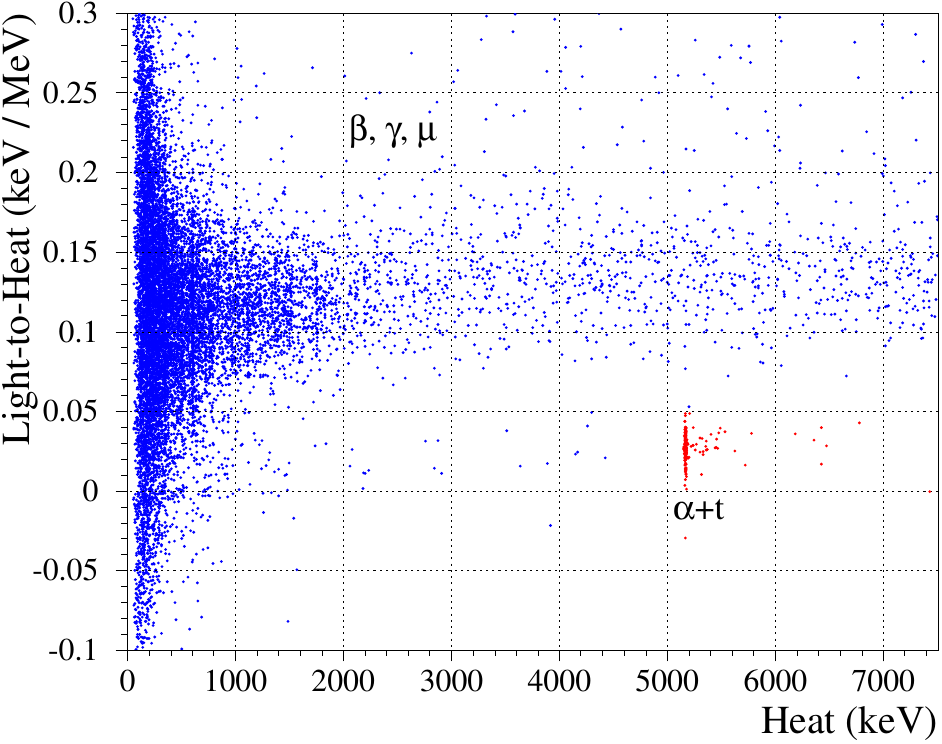}
\caption{Distribution of the Light-to-Heat parameter as a function of the Heat energy of particles detected by the Li$_2$$^{100}$MoO$_4$ scintillating bolometer (LMO-78t) in coincidences with the top (Left) and bottom (Right) LDs based on square- and circular-shaped Ge wafers, respectively. (Details of the detected scintillation signals are discussed in text.)}
\label{fig:LMO78UV_LY-vs-Heat}
\end{figure}

Examples of particle identification capability of the studied scintillating bolometers are shown in figure \ref{fig:LMO78UV_LY-vs-Heat}, where $\gamma$($\beta$)- and muon-induced events are clearly separated from $\alpha$ particles mainly represented by products of the $^6$Li(n,t)$\alpha$ reaction. Studying the coincidences between LMOs and their top LDs, we observed a notable thermal cross-talk affecting the amplitude and shape of LD signals, thus resulting to a fake estimate of the Light-to-Heat ratio, as illustrated in figure \ref{fig:LMO78UV_LY-vs-Heat} (Left). This is due to the strong pressure applied on Ge wafers by the Nylon screws towards the LMO surface. Despite this issue, the thermal cross-talk does not spoil particle identification once the noise of the photodetector is at an acceptable level. In the case of the bottom LDs having a smaller area, the frame-induced shadowing and a longer distance to the corresponding crystals, we detected very low scintillation, around 0.01\% of the heat energy measured by the LMOs. The use of the Neganov-Trofimov-Luke mode for a bottom LD (namely, the cLD-78) allowed us to get enhanced performance and thus efficient particle identification even for the case of a tiny scintillation signal and not an exceptional initial performance of the LD in absence of the Neganov-Trofimov-Luke effect.

%--------------------------------------------------
\subsubsection{Other \emph{Thick} and \emph{Slim} structure prototypes}
\label{sec:other_tests_ijclab}

After the first extensive study of the detector structure described in the previous section, we carried out a subsequent series of shorter low-temperature tests of \emph{Thick} and then \emph{Slim} design prototypes in the aboveground laboratory aiming to get a fast feedback on a further tuning of the structure components. We did not pay a lot of attention on the optimization of noise conditions and detector performance, considering a planned characterization in an underground laboratory under better background conditions, as described in the next section. Therefore, we omit details on detector performance achieved, but we provide only a short description of the studied modifications.

We investigated several key configurations towards the development of the detector structure for CROSS: 
\begin{itemize}
    \item (a) A \emph{Thick} design module with the spacers for LDs made of a 0.25-mm-thick Teflon foil (3 layers of the foil were used between an LMO crystal and a Ge wafer, while a single layer was used above it), pressed with the Nylon screws with the slit; 
    
    \item (b) A \emph{Thick} prototype, in which the LDs were clamped with the big 3D-printed clamps and squeezed with copper screws; 
    
    \item (c) An upgraded \emph{Thick} module in which we removed a few screws and pressed with the frame to reach a threshold where the curved edge of the newly 3D-printed clamps touches the top of the LD; 
        
    \item (d) The first prototype of the \emph{Slim} concept with the 3D-printed clamps; 
     
    \item (e) A \emph{Slim} design module with all spacers, including clamps, made of PTFE;  
        
    \item (f) A \emph{Slim} prototype with the 3D-printed PLA spacers used, including clamps;   

    \item (g) Several \emph{Slim} modules with upgraded 3D-printed clamps.    
\end{itemize}

After the test of the detector configurations (a)--(c), we defined the final design of the \emph{Thick} structure components for the construction of a 6-LMO-crystal array, which was then operated and studied in an underground laboratory (as detailed in section \ref{sec:tests_lsc}). In a similar way, we pre-validated a configuration of the \emph{Slim} design, (d)--(f), and used it for the construction of a 4-LMO-crystal module, tested underground with a \emph{Thick} module (see section \ref{sec:tests_lsc}). Both \emph{Thick} and \emph{Slim} designs have been successfully validated by the results of underground measurements. However, we found that performance of light detectors in terms of noise level are not as good as we expected, and we assumed it is due to insufficient clamping of the Ge wafer. Therefore, we realized a thicker version of the 3D-printed PLA spacers and validated its positive impact on the detector noise in aboveground (g) and recently in underground \cite{CUPIDalternativeStructure:2024} measurements with the \emph{Slim} type modules. Details on the first characterization of \emph{Thick} and \emph{Slim} detector structures in an underground low-background cryogenic facility are presented in the next section.

%-------------------------------------------------------------------------------------
\subsection{Underground operation of 6-crystal arrays}
\label{sec:tests_lsc}

%--------------------------------------------------
\subsubsection{Experiments in the Canfranc laboratory}

\begin{figure}[hbt]
\centering
\includegraphics[width=0.75\textwidth]{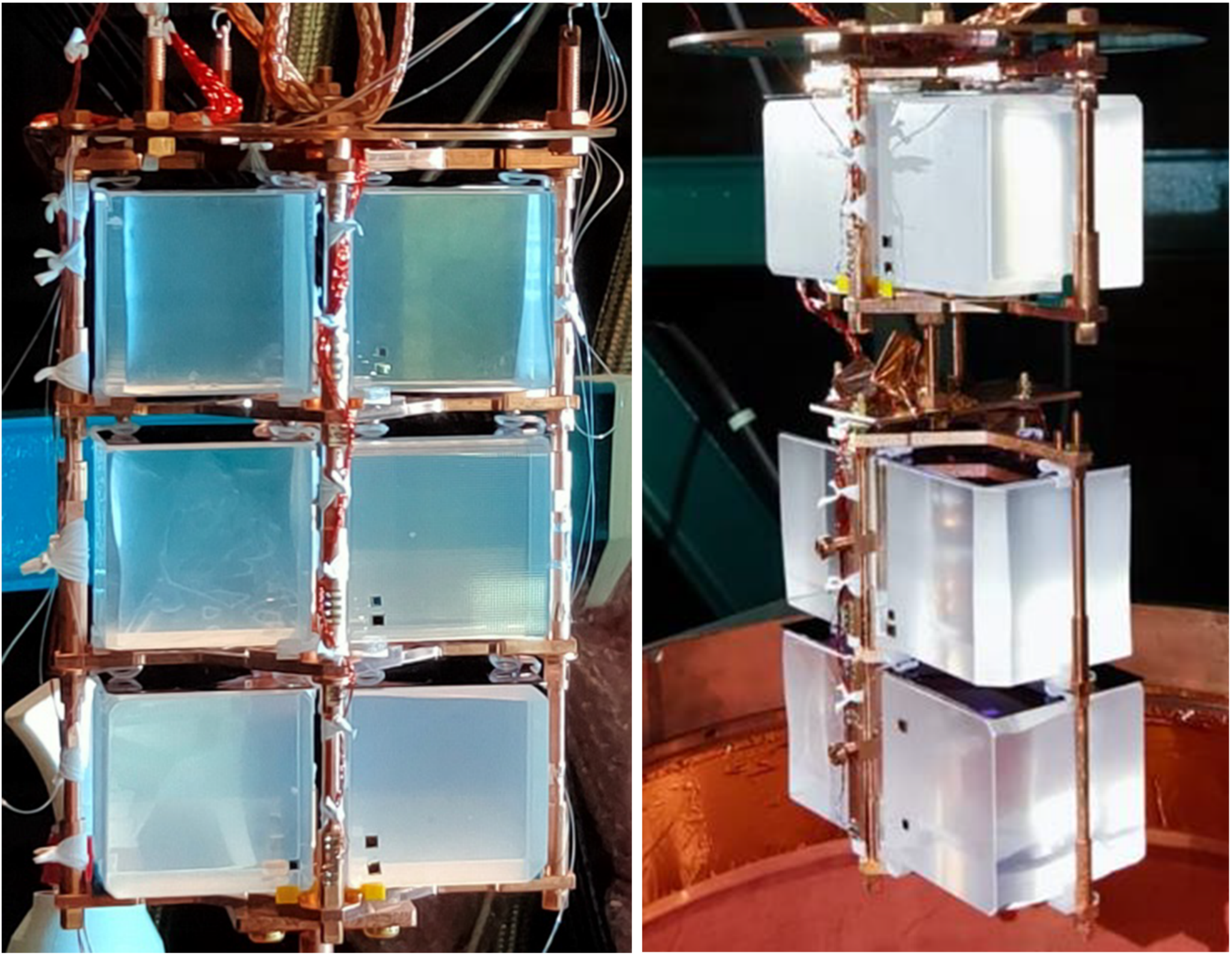}
\caption{The 6-crystal array of LMO scintillating bolometers assembled following either the \emph{Thick} structure design (Left) or using both \emph{Thick} and \emph{Slim} types simultaneously (Right). Each array was spring-suspended from the detector plate of the CROSS cryostat in the Canfranc underground laboratory (Spain).}
\label{fig:Detector_lsc}
\end{figure}

The \emph{Thick} design represented by a 6-LMO-crystal array was tested first in underground conditions (refereed here to Run~I), provided by the LSC with a rock overburden of 2450 m of water equivalent, helpful for the suppression of muon-induced background \cite{Trzaska:2019}. In two following cryogenic runs, (Run II and Run III), we operated an array with the same number of LMO crystals, but only a single 2-crystal module adopted the \emph{Thick} design, while the remaining part of the tower was constructed according to the \emph{Slim} concept. Both detector towers are shown in figure \ref{fig:Detector_lsc}. It is worth noting that the 2-crystal module of the \emph{Thick} structure was previously tested above ground (e.g. see section \ref{sec:Prototype_Thick}) and included in each tower with no change, as a reference.

%\begin{figure}[hbt]

Each array contained the LMO crystals produced from isotopically-modified molybdenum: most of them are enriched in $^{100}$Mo ($\sim$98\%) \cite{Bandac:2020,Armatol:2021b}, while a single crystal in the first tower and two samples in the second array were produced from $^{100}$Mo-depleted molybdenum \cite{Grigorieva:2020}, proven to have similar bolometric performance and radiopurity as $^{100}$Mo-enriched detectors \cite{CROSSdeplLMO:2023}. A single LMO detector of Run I was covered by a thin metal film (in the form of a grid) to test the CROSS technology of near-surface-event identification \cite{Bandac:2020,Bandac:2021} and is not considered in the present work. 
Each LMO sample was equipped with a Ge NTD (3 $\times$ 3 $\times$ 1~mm) coupled with UV-cured glue to the crystal surface; the only exception regards one sample (LMO-depl-1, reused from previous measurements \cite{CROSSdeplLMO:2023}) where the coupling is done by Araldite\textregistered ~bi-component epoxy. This epoxy has been also used to glue the heaters on the LMO surface and smaller NTDs (0.6 $\times$ 3 $\times$ 1~mm) on the Ge slabs (45 $\times$ 45 $\times$ 0.3~mm), to construct the bolometric photodetectors. The Ge wafers are coated on both sides with a thin (70 nm) SiO layer to improve light absorption. The LMO and Ge absorbers were then assembled in the Cu frame, according to the chosen \emph{Thick} or \emph{Slim} designs, using PLA 3D-printed pieces to hold and decouple the detector elements. The radioactivity of the PLA material was tested and measured to be very low. Finally, the NTDs and heaters are wire-bonded to gold-plated Kapton pads, providing electrical connection (through the set of wires) to the room-temperature electronics.

The 6-LMO-crystal arrays were operated one by one in the CROSS cryogenic set-up (C2U), described in details in \cite{Olivieri:2020,CrossCupidTower:2023a}. The C2U facility is equipped with a dilution refrigerator by CryoConcept (France), which utilises a pulse tube (Cryomech PT415) for the cooling down to 4 K. Taking into account that a pulse tube is a source of vibrations, that can affect detectors performance \cite{Olivieri:2017}, the cryostat is assisted by the Ultra-Quiet Technology{\texttrademark} \cite{UQT} to decouple mechanically the pulse tube from the dilution unit. In order to further reduce residual pulse-tube-injected vibrations, a 3-spring-based mechanical suspension of the tower was used in Run I. After the completion of Run I and the opening of the cryostat, we observed that the suspension of the tower exhibited a too strong mechanical coupling through the flexible thermalization (Cu braids), explaining the noise issue faced in these measurements and detailed below. In the next experiment (Run II) we used a single-spring suspension and a longer thermal link. In the last measurement (Run III) we tested the same tower (Run II) with a new detector suspension system developed for the CROSS experiment, consisting of a set of three springs installed at the 1 K stage provided with magnetic dampers (the natural frequency of the system is 3.5 Hz) \cite{CROSS_Magnetic_dampers:2023}. 

The experimental volume of the cryostat is protected laterally and at the bottom by a 25-cm-thick Pb shield, and by a 13-cm-thick Pb and Cu internal shielding from the top part of the cryostat. Low-radioactivity lead (but with a remarkable $^{210}$Pb content \cite{CROSSdeplLMO:2023}) has been used in all parts of the shielding, except for the internal one where the $^{210}$Pb content is low. In addition, an anti-radon box is installed around the shield and is continuously flushed with deradonized air ($\sim$1 mBq/m$^3$ of Rn \cite{PerezPerez:2022}). A muon veto system has been installed as well, surrounding the cryostat, inside the hut.

Each channel was read out using a room-temperature low-noise DC electronics, reconditioned from the boards of the Cuoricino experiment \cite{Arnaboldi:2002}. The DAQ was composed of two 12-channel boards with an integrated 24-bit ADC \cite{Carniti:2020,Carniti:2023}. A programmable 6-pole Bessel-Thomson anti-aliasing filter was also used in the detector readout (a low-pass filter cut-off frequency was set at 300~Hz). We used the heaters glued at the LMO crystals to inject thermal pulses into them using a wave-function generator (Keysight 33500B), thus mimicking particle-induced response. Heater-induced pulses can be used for detector optimization and/or stabilization. For the light detectors, we used a room-temperature LED (880 nm emission maximum) instead of heaters to mimic the device response, flushing the experimental volume with a burst of photons transmitted via an optic fiber. In addition, a removable $^{232}$Th $\gamma$ source was inserted periodically inside the Pb shield for calibration of both LMO and LD bolometers.

\begin{table}
\centering
\caption{Low-temperature experiments in the CROSS cryogenic facility at the LSC underground laboratory carried out to test 6-LMO-crystal arrays using the \emph{Thick} and/or \emph{Slim} detector structures. We are quoting: a period of data taking; its duration; a corresponding duty cycle of the facility; and the type of measurement ($^{232}$Th calibration / background) for the acquired data.}
\smallskip
\begin{tabular}{cccccc}
\hline
Run & Detector & Data taking  & Duration & Duty  & Data type \\
~  & structure & period & (d) & cycle   &  (\% of total data) \\
\hline
\hline
I   & \emph{Thick}                & Jun. 2021 -- Oct. 2021 & 126 & 75\%  & Th (77\%), Bkg (23\%)    \\
II  & \emph{Thick} \& \emph{Slim} & Jan. 2022 -- Apr. 2022 & 100 & 92\%  & Th (65\%), Bkg (35\%)  \\
III & \emph{Thick} \& \emph{Slim} & Aug. 2022 -- Oct. 2022 &  82 & 95\%  & Th (52\%), Bkg (35\%)    \\

\hline
\end{tabular}
\label{tab:Runs_LSC}
\end{table}

Some details about the carried out experiments are given in table \ref{tab:Runs_LSC}. The detector arrays were operated at similar temperatures of the heat sink, 15--16~mK, regulated on the detector floating plate. The best working point of each detector channel, defined as the highest $S/N$ ratio, was determined using heater (for LMOs) or LED (for LDs) induced signals. Depending on the experimental conditions, detectors were biased with NTD currents between hundreds pA and a few nA; the corresponding NTD resistances were measured to lie between a few M$\Omega$ and about a hundred M$\Omega$.   
Continuous data of all channels were acquired simultaneously with a sampling frequency of 2~kS/s. A fast low-level analysis was done during the data taking using our online monitor of the experimental conditions. The routine analysis was done off-line using the data processing tools described in section \ref{fig:LD_gravity_LY}. 
The energy threshold was varied between 5 RMS and 10 RMS of the filtered baseline noise, depending on the detector counting rate (i.e. the absence / presence of the $^{232}$Th source).

%--------------------------------------------------
\subsubsection{Characterization of LMO thermal detectors}

As the target application of the tested low-temperature detectors is double-beta decay search, we characterize LMO bolometers in terms of the following parameters: 
a) the detector sensitivity, defined as a voltage signal per unit of deposited energy; 
b) the energy resolution of the baseline noise traces after the optimum filtering; 
c) the energy resolution of a $\gamma$-ray peak in the proximity of the $^{100}$Mo ROI, induced by 2615 keV $\gamma$ quanta of $^{208}$Tl during the LMO irradiation with a $^{232}$Th source. 

In Run I, we observed that the LMO detectors were impacted by a significant noise excess, in particular in the frequency interval of 10--50 Hz, in comparison to the previous measurement in the C2U set-up with the 12-crystal array of LMO scintillating bolometers \cite{CrossCupidTower:2023a}. Therefore, despite the good sensitivities demonstrated by the LMO bolometers (30--90 nV/keV) as for particle thermal detectors of such type and size and for the observed range of NTD resistances, the baseline noise resolution of the LMO thermal detectors (2--10 keV FWHM) was drastically worse with respect to the expectations for most of them. Indeed, our ``reference'' detectors exhibit a factor 3--5 larger values of the baseline noise resolution compared to the performance achieved in the aboveground laboratory. Despite large values of the baseline noise, the energy resolution measured with the 2615~keV $\gamma$ quanta of $^{208}$Tl is rather high (6--7 keV FWHM) for detectors less affected by noise and is still good ($\sim$10~keV FWHM) even for the noisiest bolometers. This observation shows that the high noise level measured at the baseline has scarce impact on the resolution at high energy. It can be attributed to noise pick-ups in the signal readout.

\begin{figure}[hbt]
\centering
\includegraphics[width=0.7\textwidth]{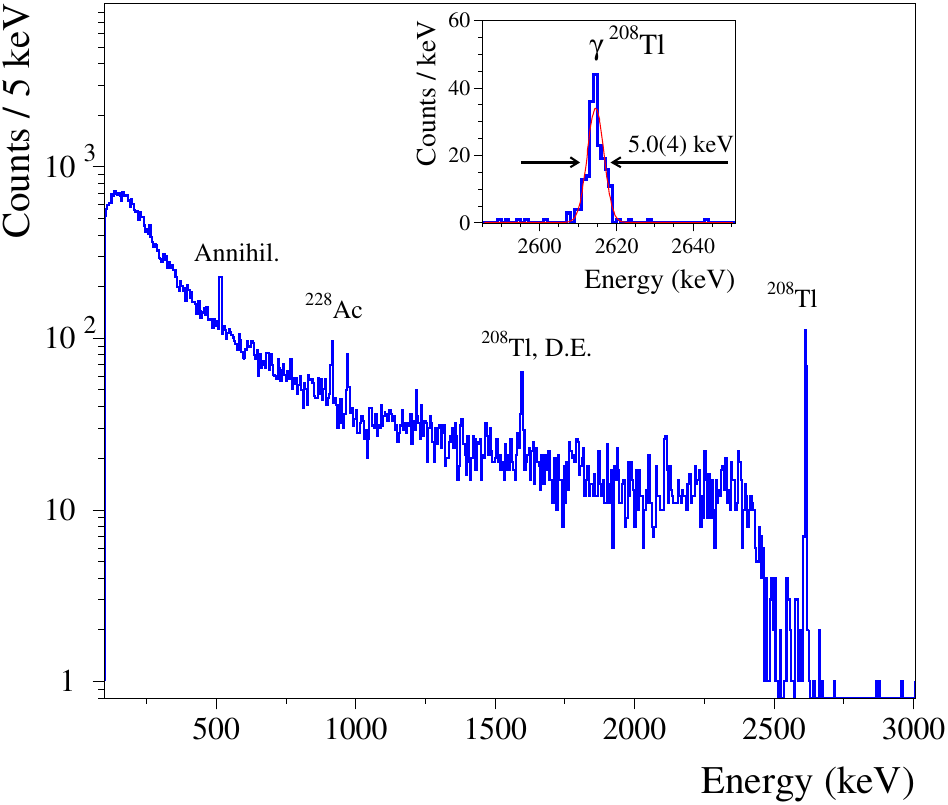}
\caption{Energy spectrum of a $^{232}$Th source measured with one of the best-performing $^{100}$Mo-enriched Li$_2$MoO$_4$ bolometers (LMO-78b) tested at the LSC. The most intense $\gamma$-ray peaks of the spectrum are labeled (D.E. is the double escape peak from 2615 keV photons). The inset shows the 2615-keV $\gamma$ peak from $^{208}$Tl and a fit with a Gaussian function; the energy resolution is 5.0(4) keV FWHM.}
  \label{fig:LMO_Th_spectrum}
\end{figure}

Thanks to the proper functionality of the tower suspension in Run II, a notable improvement in the baseline noise resolution has been demonstrated by all the LMO bolometers (0.8--1.7 keV FWHM), except a single module (LMO-55t) showing modest progress. Also, the chosen working points are characterized by a factor 2--3 higher sensitivities (80--300 nV/keV). The energy resolution of the 2615 keV $\gamma$ peak is measured as 5--7 keV FWHM for all, but $\sim$10 keV FWHM for the detector (LMO-55t) with the noise excess. An example of the energy spectrum measured by an LMO bolometer in a calibration with the $^{232}$Th source of $\gamma$ quanta is shown in figure \ref{fig:LMO_Th_spectrum}.

LMO performance in Run III is similar to that in Run II in terms of the achieved baseline noise resolution (1--6 keV FWHM), but the sensitivities have been reduced (50--150 nV/keV), in correlation with the decreased NTD resistances. In addition to LMO-55t, LMO-depl-2 became more affected by noise too, which resulted to worsening the energy resolution of these detectors. Other bolometers show good energy resolution measured with the 2615 keV $\gamma$s (5--7 keV FWHM).

\begin{figure}[hbt]
\centering
\includegraphics[width=0.49\textwidth]{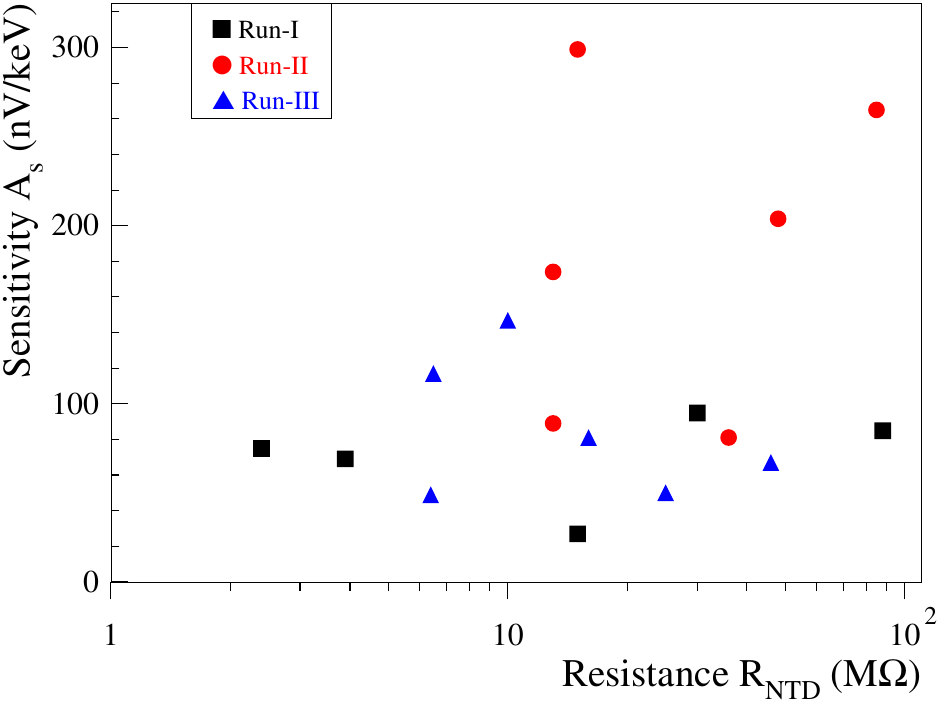}
\includegraphics[width=0.49\textwidth]{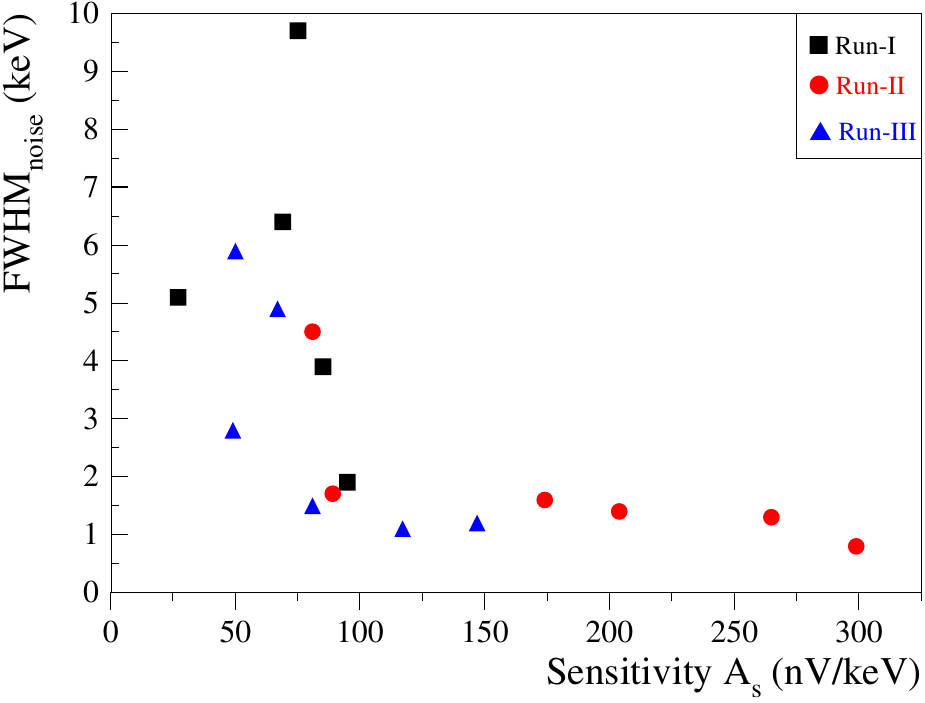}
\caption{Parameters (NTD resistance, sensitivity and noise energy resolution) of Li$_2$MoO$_4$-based bolometers operated in the CROSS facility at the LSC (Spain) over three runs.}
  \label{fig:LMO_performance_lsc}
\end{figure}

\begin{figure}[hbt]
\centering
\includegraphics[width=0.65\textwidth]{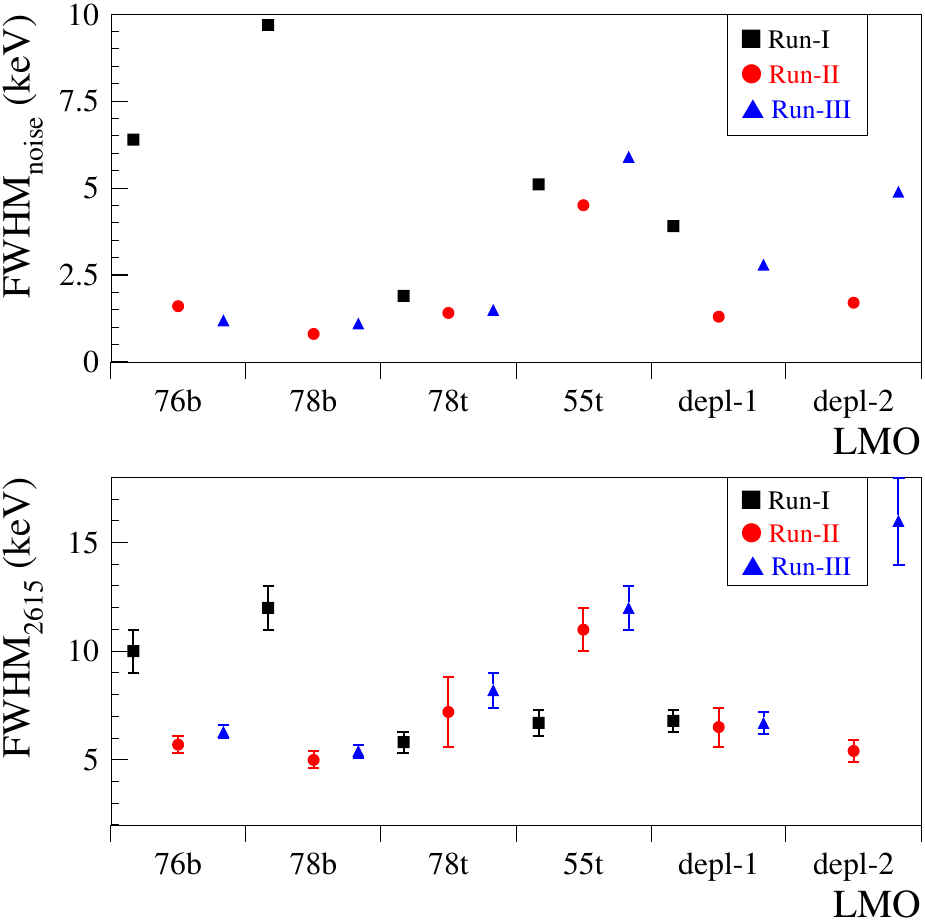}
\caption{Energy resolution of Li$_2$MoO$_4$-based bolometers measured for noise samples (top panel) and at 2615-keV $\gamma$ peak (bottom panel) in three cryogenic runs in the CROSS underground facility.}
  \label{fig:LMO_FWHM_lsc}
\end{figure}

The best performances achieved by the LMO bolometers in all three underground experiments are summarized in table \ref{tab:LMO_performance} and illustrated in figures \ref{fig:LMO_performance_lsc} and \ref{fig:LMO_FWHM_lsc}. 
By comparing results of all the runs in terms of achieved energy resolution (figure \ref{fig:LMO_FWHM_lsc}), we can conclude that LMO bolometers being fabricated using either \emph{Thick} or \emph{Slim} design of the detector structure and operated in a pulse-tube cryostat are capable to reach high performance, compatible with CROSS demands for high-sensitivity searches for $0\nu\beta\beta$ decay of $^{100}$Mo. In addition, we see that some detector readout channels of the C2U set-up can have a noise excess despite the advanced damping system of the detector plate installed recently inside the cryostat.

\begin{table}
\centering
\caption{Best performance of the 6-LMO-crystal arrays, considering both \emph{Thick} and \emph{Slim} designs, operated in a pulse-tube cryostat at the LSC over three consecutive cryogenic runs. For each bolometer, we quote the working point parameters (NTD resistance $R_{NTD}$ at a given current $I_{NTD}$), the detector sensitivity ($A_s$), the LMO energy resolution (the full width at the half maximum) measured for zero energy deposition (FWHM$_{noise}$) and for 2615 keV $\gamma$ of $^{208}$Tl (FWHM$_{2615}$).}
\smallskip
\begin{tabular}{lccccccc}
\hline
Channel  & Run & Detector & $R_{NTD}$ & $I_{NTD}$ & $A_s$ & FWHM$_{noise}$ & FWHM$_{2615}$  \\
~ & ID  & structure & (M$\Omega$) & (nA)   & (nV/keV) & (keV) & (keV)  \\
\hline
\hline
LMO-76b    & I   & \emph{Thick} & 3.9 & 2.6  & 69  & 6.4  & 10 $\pm$ 1     \\
~          & II  & \emph{Thick} & 13  & 1.0  & 174 & 1.6 & 5.7 $\pm$ 0.4    \\
~          & III & \emph{Thick} & 10  & 1.0  & 147 & 1.2 & 6.3 $\pm$ 0.3    \\
\cline{1-8}
LMO-78b    & I   & \emph{Thick} & 2.4 & 2.6  & 75  & 9.7  & 12 $\pm$ 1     \\
~          & II  & \emph{Thick} & 15  & 0.55 & 299 & 0.80 & 5.0 $\pm$ 0.4    \\
~          & III & \emph{Thick} & 6.5 & 1.0  & 117 & 1.1 & 5.4 $\pm$ 0.3   \\
\hline
LMO-78t    & I   & \emph{Thick} & 30  & 0.55 & 95  & 1.9 & 5.8 $\pm$ 0.5      \\
~          & II  & \emph{Slim} & 48  & 0.35 & 204 & 1.4 & 7.2 $\pm$ 1.6   \\
~          & III & \emph{Slim}  & 16  & 1.0  & 81  & 1.5 & 8.2 $\pm$ 0.8    \\
\cline{1-8}
LMO-55t    & I   & \emph{Thick} & 15  & 0.70 & 27  & 5.1  & 6.7 $\pm$ 0.6      \\
~          & II  & \emph{Slim} & 36  & 0.36 & 81  & 4.5  & 11 $\pm$ 1    \\
~          & III & \emph{Slim}  & 25  & 0.36 & 50  & 5.9  & 12 $\pm$ 1     \\
\cline{1-8}
LMO-depl-1 & I   & \emph{Thick} & 88  & 0.18 & 85  & 3.9  & 6.8 $\pm$ 0.5      \\
~          & II  & \emph{Slim} & 85  & 0.18 & 265 & 1.3  & 6.5 $\pm$ 0.9    \\
~          & III & \emph{Slim}  & 6.4 & 1.0  & 49  & 2.8  &  6.7 $\pm$ 0.5   \\
\cline{1-8}
LMO-depl-2 & I   & \emph{Thick} & n.a.& n.a. & n.a.& n.a. & n.a.      \\
~          & II  & \emph{Slim} & 13  & 0.55 & 89  & 1.7  & 5.4 $\pm$ 0.5      \\
~          & III & \emph{Slim}  & 46  & 0.36 & 67  & 4.9  & 16 $\pm$ 2      \\
\hline
\end{tabular}
\label{tab:LMO_performance}
\end{table}

%----------------------------------------------------------------------------------
\subsubsection{Performance of bolometric Ge light detectors}

The Ge bolometers used in the towers are characterized in the same way as LMO detectors, except for a study of the energy resolution with high-energy $\gamma$ quanta taking into account the small volume of these thermal detectors. It is worth noting that the primary goal of the used Ge bolometers is the detection of scintillation light, which is typically a small fraction ($\sim$0.1\%--1\% depending on the scintillation material \cite{Poda:2021}) of the energy release induced by the particle interaction inside a scintillating bolometer. Consequently, the energy resolution of such devices is limited by poor photon statistics as well as by its energy dependence; thus, it plays a secondary role in the considered application. On the contrary, the baseline noise resolution becomes the crucial parameter, demonstrating the capability of particle identification.

\begin{table}
\centering
\caption{Best performance of bolometric Ge light detectors used in the 6-LMO-crystal arrays, \emph{Thick} and \emph{Slim} designs, and operated in a pulse-tube cryostat at the LSC over three consecutive cryogenic runs. For each operational LD we are quoting the working point parameters (NTD resistance $R_{NTD}$ at a given current $I_{NTD}$), the detector sensitivity ($A_s$), and the baseline noise resolution after optimum filtering (RMS$_{noise}$).}
\smallskip
\begin{tabular}{lcccccc}
\hline
Channel  & Run & Detector & $R_{NTD}$ & $I_{NTD}$ & $A_s$ & RMS$_{noise}$  \\
~ & ID  & structure & (M$\Omega$) & (nA)   & ($\mu$V/keV) & (keV) \\
\hline
\hline
sLD-76b    & I   & \emph{Thick} & 3.8 & 2.0  &  0.76  & 0.99    \\
~          & II  & \emph{Thick} & 32  & 0.55 & 2.3  & 0.059    \\
~          & III & \emph{Thick} & 39  & 0.18 & 2.9  & 0.088   \\
\cline{1-7}
sLD-78b    & I   & \emph{Thick} & 3.5 & 2.6  & 1.5  & 0.56    \\
~          & II  & \emph{Thick} & 11  & 1.0  & 2.6  & 0.068    \\
~          & III & \emph{Thick} & 10  & 1.0  & 2.5  & 0.054   \\
\hline
sLD-78t    & I   & \emph{Thick} & 0.3 & 14   &  0.32  & 0.65     \\
~          & II  & \emph{Slim} & 3.6 & 2.9  &  0.60  & 0.36    \\
~          & III & \emph{Slim}  & 10  & 1.0  &  0.70  & 0.35   \\
\cline{1-7}
sLD-55t    & I   & \emph{Thick} & 0.3 & 14   &  0.23  & 1.2     \\
~          & II  & \emph{Slim} & -   & -    & -     & -      \\
~          & III & \emph{Slim}  & 15  & 0.55 & 1.9  & 0.10    \\
\cline{1-7}
sLD-depl-1 & I   & \emph{Thick} & 0.4 & 14   &  0.31  & 0.97     \\
~          & II  & \emph{Slim} & 3.0 & 4.8  &  0.70  & 0.14    \\
~          & III & \emph{Slim}  & 14  & 1.0  & 1.5  & 0.20   \\
\cline{1-7}
sLD-depl-2 & I   & \emph{Thick} & n.a.& n.a  & n.a   & n.a    \\
~          & II  & \emph{Slim} & 120 & 0.18 & 3.0  & 0.18    \\
~          & III & \emph{Slim}  & -   & -    & -     & -       \\
\hline
\end{tabular}
\label{tab:LD_performance}
\end{table}

\begin{figure}[hbt]
\centering
\includegraphics[width=0.49\textwidth]{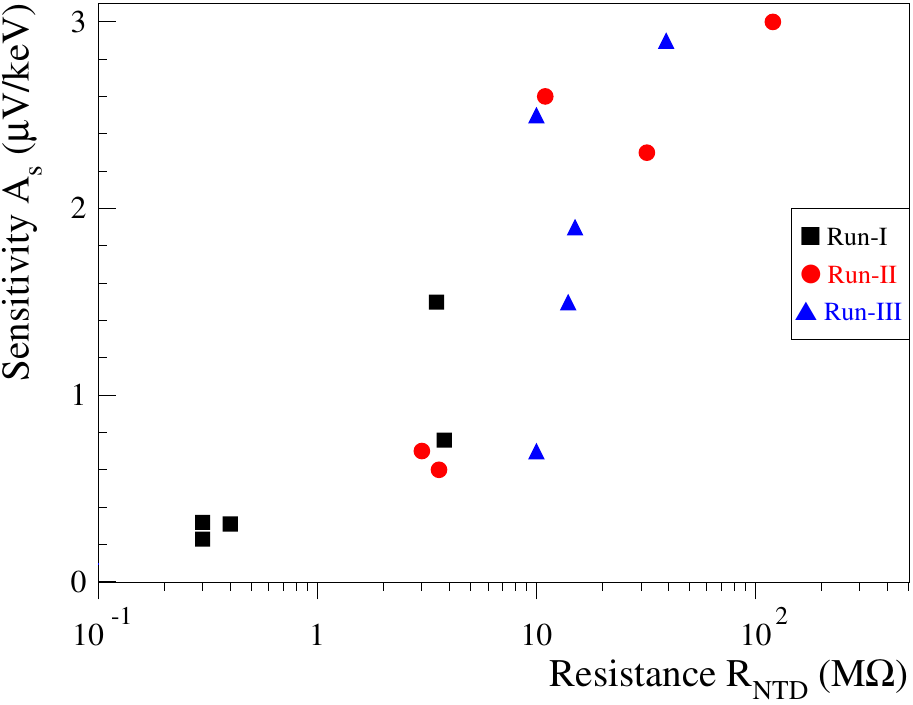}
\includegraphics[width=0.49\textwidth]{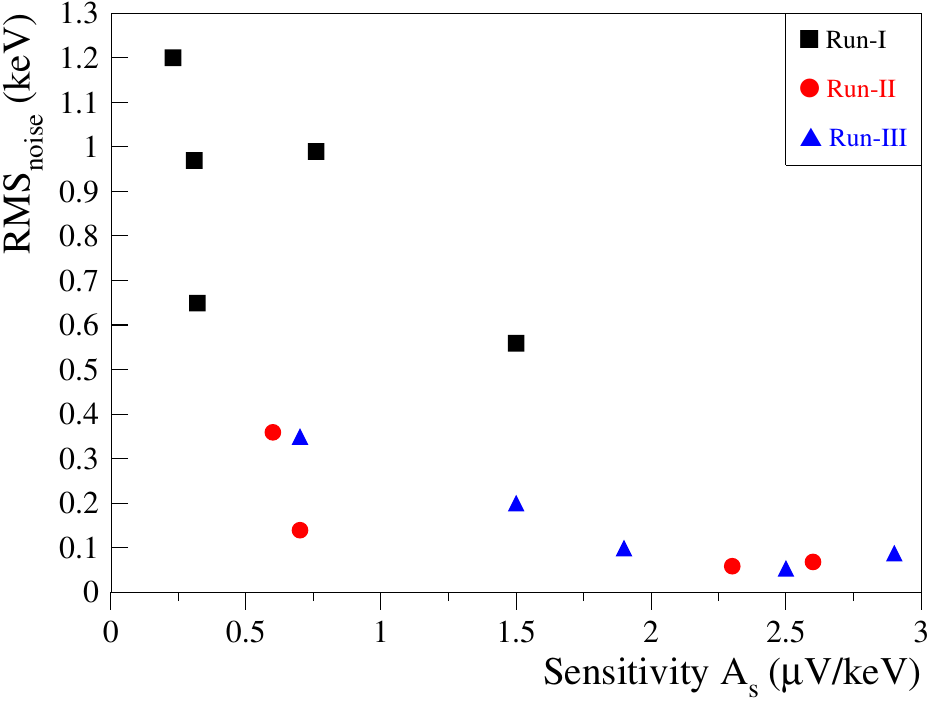}
\includegraphics[width=0.65\textwidth]{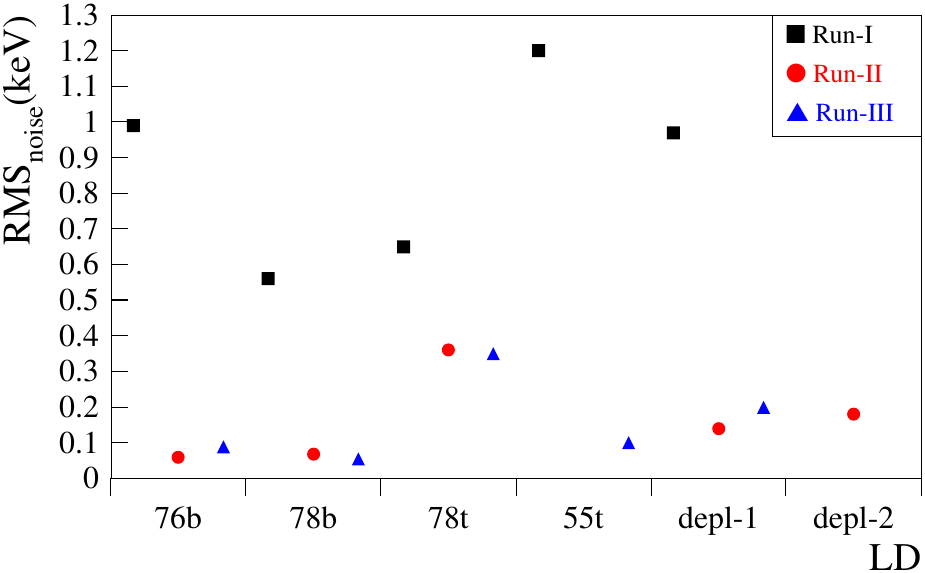}
\caption{Parameters (NTD resistance, sensitivity and noise energy resolution) of bolometric Ge light detectors characterized over three cryogenic runs in the CROSS pulse-tube-based dilution refrigerator.}
  \label{fig:LD_performance}
\end{figure}

Similarly to LMOs, the performance of LDs in Run I was strongly affected by vibration-induced noise. Therefore, we operated NTDs at high currents (2--14 nA), thus reducing the resistance of the sensors (0.3--4 M$\Omega$) and the detectors' sensitivity (0.2--1.5 $\mu$V/keV) to rather low values. The baseline noise of LDs in Run I was found to be atypically high (0.6--1.0 keV RMS). A significant improvement of the noise conditions over Run II is evident for LDs (0.06--0.4 keV RMS), which were operated at colder working temperatures resulting to higher NTD resistances (3--120 M$\Omega$) and higher sensitivities (0.6--3.0 $\mu$V/keV). In the set-up upgraded with the magnetic-damping-based suspension, Run III, the performance of LDs is found to be similar to Run II results. The best performance obtained with the bolometric Ge light detectors in all three underground Runs are listed in table \ref{tab:LD_performance} and visualized in figure \ref{fig:LD_performance}.

It is worth noting an indication that LDs assembled in the \emph{Thick} structure have overall lower noise than that of LDs in the \emph{Slim} design. In addition, a twice better noise level was achieved with LDs operated in the C2U set-up in runs \cite{Armatol:2021b,CrossCupidTower:2023a,CROSSdeplLMO:2023,CROSS_Magnetic_dampers:2023} preceding the ones described here. In these previous runs, the Ge wafers were clamped by PTFE elements inside an individual copper frame, assuring a very effective holding of the detectors. Thus we can conclude that the used version of the PLA clamps is not fully sufficient in holding tightly the Ge wafers as obtained in previous tests. To overcome this issue we slightly modified the clamps to improve the tightness of the Ge wafers in the same detector structures described here. This solution was first validated in a set of aboveground measurements (section \ref{sec:other_tests_ijclab}) and finally confirmed in a recent underground test of a 10-crystal array containing 10 Ge LDs \cite{CUPIDalternativeStructure:2024}.

%----------------------------------------------------------------------------------
\subsubsection{Particle identification with LMO scintillating bolometers}

\begin{figure}[hbt]
\centering
\includegraphics[width=0.49\textwidth]{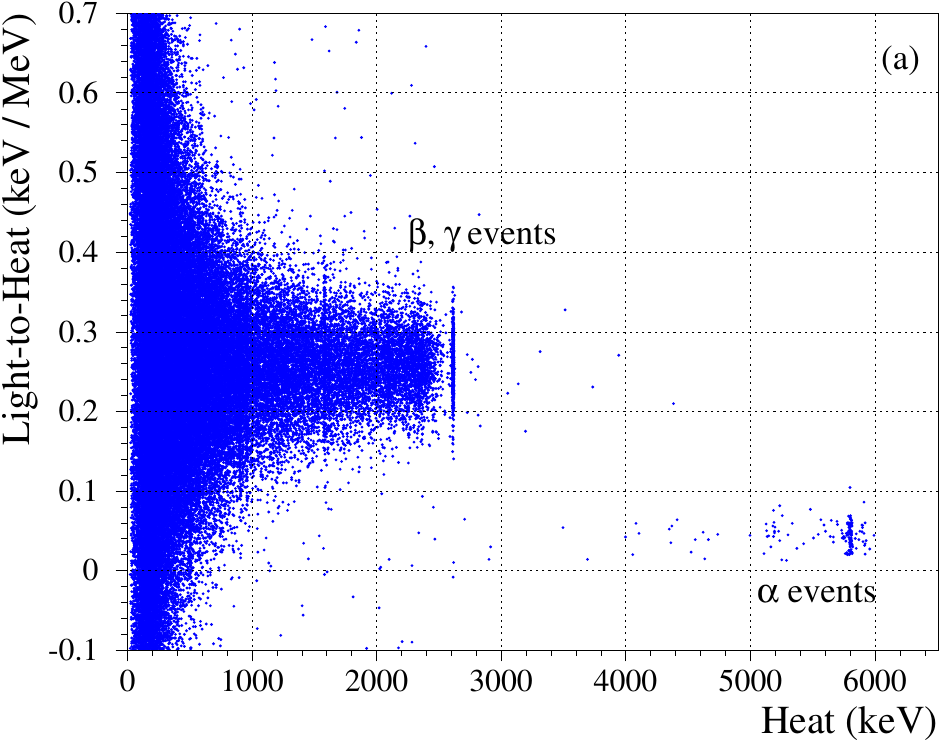}
\includegraphics[width=0.49\textwidth]{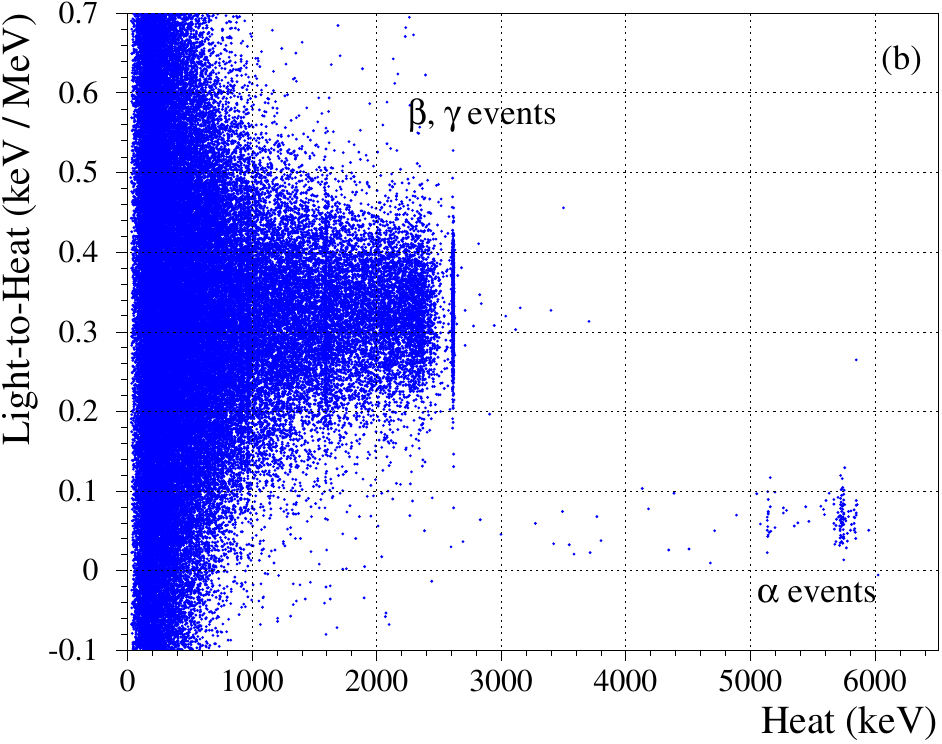}
\includegraphics[width=0.49\textwidth]{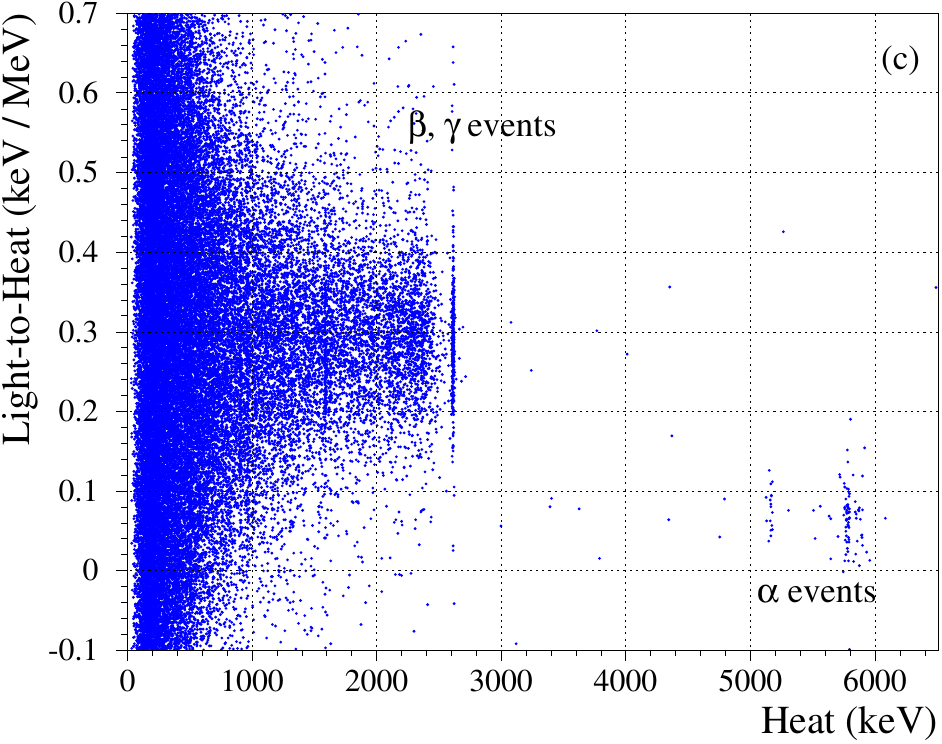}
\includegraphics[width=0.49\textwidth]{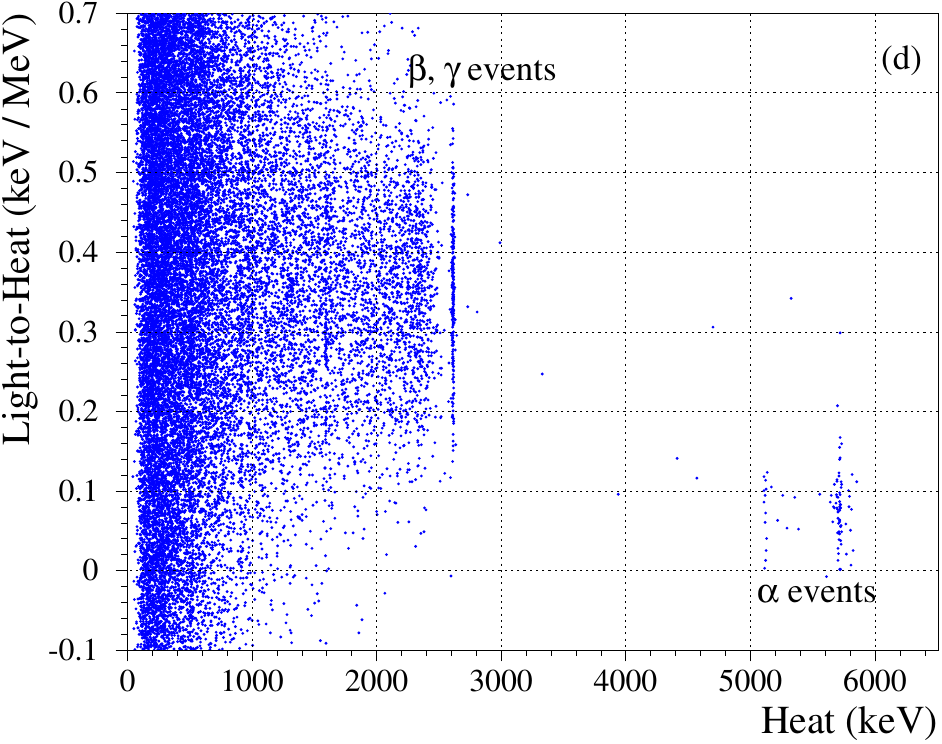}
\caption{Distribution of the Light-to-Heat parameter as a function of the Heat energy for $\gamma$($\beta$) and $\alpha$ events detected in Run III calibration measurements (163 h) with the Li$_2$MoO$_4$ scintillating bolometers LMO-78b (a), LMO-76b (b), LMO-55t (c), and LMO-depl-1 (d). The baseline noise resolution of the bolometric Ge LDs coupled to LMOs is 0.05, 0.09, 0.10, and 0.20 keV RMS, respectively.}
\label{fig:LMO_LY-vs-Heat}
\end{figure}

As mentioned above, the performance of LDs (i.e. baseline energy resolution) plays a significant role in the efficiency of scintillation-assisted particle identification of scintillating bolometers. This can be also illustrated by the results of underground measurements with LMOs, shown in figure \ref{fig:LMO_LY-vs-Heat}, demonstrating the impact of the rock shielding on detector background in comparison with the aboveground test (see figure \ref{fig:LMO78UV_LY-vs-Heat}). 

The dominant part of events in figure \ref{fig:LMO_LY-vs-Heat} belongs to the $\gamma$($\beta$) band --- with the clearly seen 2615-keV end-point of the most intense natural $\gamma$-radioactivity. 
The mean Light-to-Heat ratio for $\gamma$($\beta$) events detected by the scintillating bolometers of the present study is 0.26--0.32 keV/MeV (the harmonic mean is 0.30 keV/MeV), which is similar to the results reported recently for another open detector structures ($\sim$0.2--0.3~keV/MeV) \cite{Armatol:2021a,Alfonso:2022,CrossCupidTower:2023a}.

A few islands of events seen in the right bottom part in figure \ref{fig:LMO_LY-vs-Heat} are characterized by a high heat energy release, about 5 MeV, however their Light-to-Heat values are reduced compared to $\gamma$($\beta$)'s; this shows the quenching of scintillation light emission induced by $\alpha$ particle interaction in the LMO medium, typical for crystal scintillators, particularly operating as detectors at low temperatures \cite{Poda:2021}. Therefore, we can attribute these events to $\alpha$ particles (including a combined signal from $\alpha$'s and other types of particles emitted in the involved nuclear reactions or decays). Figure \ref{fig:LMO_LY-vs-Heat} shows that, despite the relatively low scintillation signal of $\sim$0.3 keV/MeV (as expected in the structure with no reflective cavity around the LMO crystals), $\alpha$ particles can be clearly identified and removed as potential sources of background in the $^{100}$Mo ROI, at $\sim$3 MeV, once the baseline noise of the bolometric photodetector is sufficiently low. For such light collection conditions, we expected to reach a highly-efficient $\alpha$/$\gamma$($\beta$) separation, consisting of a rejection of 99.9\% of $\alpha$s keeping about 90\% of $\gamma$($\beta$) events, using LDs characterized by a baseline noise $\sim$150 eV RMS. This estimate is in agreement with the present experimental results showing good $\alpha$/$\gamma$($\beta$) separation for all LMOs except a single crystal coupled to a LD, which exhibited a baseline noise twice higher than required. 

Despite a sufficient particle identification capability achieved in the present work, we have to admit that the noise of most LDs is still at the edge of the acceptable value, which is risky in view of a larger scale experiment. Therefore, we consider to use the upgraded version of LDs, i.e. Ge wafers with Al electrodes deposited on one surface, to enhance the detector performance exploiting the Neganov-Trofimov-Luke effect. Indeed, the applied bias on the Al electrode allows for the collection of the freed charge carriers, inducing a heat release proportional to the applied voltage. Prototypes of such detectors, particularly tested in the present work (see section \ref{sec:Prototype_Thick}), show the possibility to get a factor $\sim$10 gain in the detected signal by applying just tens of volts on the electrodes, which would allow us to drastically increase the $S/N$ ratio of LDs, thus securing particle identification. Moreover, the increased $S/N$ ratio can be exploited for more efficient rejection of the random coincidences of $2\nu\beta\beta$ events, which are considered as one of the major sources of background in next-generation bolometric experiments with $^{100}$Mo.

%==============================================================================
\section{Conclusions}

We report on the development of a mechanical structure for bolometric detector arrays in the CROSS experiment, which could also satisfy the demands of future experiments to search for neutrinoless double-beta decay in $^{100}$Mo, like CUPID. The structure follows a modular principle allowing us to assemble detector towers containing two modules of scintillating bolometers per floor. The individual detection module is based on a cubic crystal with a 45-mm side and a thin square-shaped Ge wafer acting as a photodetector. We developed two designs of the structure with a different ratio of the major passive element (copper) to the crystal mass (lithium molybdate, LMO): a) \emph{Thick} design --- a more rigid version having around 15\% of Cu over LMO mass; b) \emph{Slim} --- a fine version of the structure containing only $\sim$6\% of Cu/LMO mass. 

We validated both designs of the detector structure in several low-temperature measurements carried out at aboveground (IJCLab, France) and underground (LSC, Spain) laboratories. In particular, we exploited the cryogenic facility of the CROSS experiment in LSC to test a 6-crystal array (\emph{Thick} and mixed designs) over three subsequent cryogenic measurement campaigns (runs). We observed that despite not fully optimized noise conditions of the set-up, all LMO bolometers were capable to reach high energy resolution, 5--7~keV FWHM at 2615 keV $\gamma$ quanta, close to the region of interest for $^{100}$Mo (3034 keV). Taking into account the low scintillation light yield of the LMO material and the absence of a reflective cavity around the crystals, the scintillation signal measured by bolometric Ge light detectors is rather low, around 0.3 keV (150 photons) per 1~MeV heat energy deposited in LMOs facing the bottom side of LDs. 
Despite the low efficiency of scintillation light collection in the open detector structure, the performance of most light detectors (with a baseline noise of $\sim$50--200 eV RMS) is compatible with efficient $\alpha$/$\gamma$($\beta$) separation based on the particle-dependent difference in the amount of scintillation light emitted by LMOs (a factor 5 in favour of $\gamma$($\beta$)'s with respect to $\alpha$'s). To ensure a highly-efficient particle identification (even for photodetectors with noise excess) and to improve the capability of light detectors for the rejection of random coincidences, we plan to use upgraded bolometric light detectors, aiming at enhancing their performance via the thermal signal amplification achieved through the Neganov-Trofimov-Luke effect.

%==============================================================================
\acknowledgments

This work is supported by the European Commission (Project CROSS, Grant No. ERC-2016-ADG, ID 742345) and by the Agence Nationale de la Recherche (ANR France; Project CUPID-1, ANR-21-CE31-0014). We acknowledge also the support of the P2IO LabEx (ANR-10-LABX0038) in the framework ``Investissements d'Avenir'' (ANR-11-IDEX-0003-01 -- Project ``BSM-nu'') managed by ANR, France. 
This work was also supported by the National Research Foundation of Ukraine under Grant No. 2023.03/0213 and by the National Academy of Sciences of Ukraine in the framework of the project ``Development of bolometric experiments for the search for double beta decay'', the grant number 0121U111684. 
Russian and Ukrainian scientists have given and give crucial contributions to CROSS. For this reason, the CROSS collaboration is particularly sensitive to the current situation in Ukraine. The position of the collaboration leadership on this matter, approved by majority, is expressed at \href{https://a2c.ijclab.in2p3.fr/en/a2c-home-en/assd-home-en/assd-cross/}{https://a2c.ijclab.in2p3.fr/en/a2c-home-en/assd-home-en/assd-cross/}.

%%\bibliographystyle{plain}
%\bibliographystyle{JHEP}
%\bibliography{Bibliography}

\end{document}